\documentclass[journal]{IEEEtran}

\usepackage{cite}
\usepackage{amsmath,amssymb,amsfonts}
\usepackage{algorithm}
\usepackage{algpseudocode}
\usepackage{graphicx}
\usepackage{textcomp}
\usepackage{tabularx}
\usepackage{booktabs}
\usepackage{makecell}
\usepackage{bm}

\newcommand{\bvec}[1]{\bm{#1}}

\def\BibTeX{{\rm B\kern-.05em{\sc i\kern-.025em b}\kern-.08em
    T\kern-.1667em\lower.7ex\hbox{E}\kern-.125emX}}

\begin{document}

\title{TP-CRIV: A Framework for Third-Party Challenge-Response Identity Verification of AI Models}

\author{Teruki Sano, Minoru Kuribayashi, Masao Sakai, Shuji Isobe, Eisuke Koizumi, Zhang Zhang, and Satoru Matsumoto%
\thanks{Teruki Sano is with the Graduate School of Information Sciences, Tohoku University, Japan.}%
\thanks{Minoru Kuribayashi, Masao Sakai, Shuji Isobe, Eisuke Koizumi, Zhang Zhang, and Satoru Matsumoto are with the Center for Data-driven Science and Artificial Intelligence, Tohoku University, Japan.}%
\thanks{Corresponding author: Teruki Sano (e-mail: sano.teruki.r2@dc.tohoku.ac.jp).}%
\thanks{This study was supported by the JSPS KAKENHI (25K15225, 26H02540) and Support Center for Advanced Telecommunications Technology Research (SCAT), Japan.}%
\thanks{This work has been submitted to the IEEE for possible publication. Copyright may be transferred without notice, after which this version may no longer be accessible.}%
}

\maketitle

\begin{abstract}
Artificial intelligence (AI) models are
increasingly deployed through remote services,
making model misappropriation a growing concern.
Existing approaches, including watermarking,
fingerprinting, and model similarity analysis,
primarily rely on predefined evidence or direct
behavioral comparison and do not explicitly evaluate
whether the claimant currently possesses and can
utilize model-dependent information relevant to the
claimed model identity.

In this paper, we propose Third-Party Challenge-Response Identity Verification (TP-CRIV) for AI models. TP-CRIV targets a third-party verification setting in which the verifier has neither white-box nor API access to the claimant's model, can interact with the suspicious deployed service only through its ordinary black-box inference interface, and does not require protocol-specific cooperation from the service provider. Under these constraints, the framework enables the verifier to obtain empirical evidence as to whether the claimant locally possesses a model satisfying a predeclared identity relative to the deployed model. Verification is conducted under fresh, previously undisclosed requirements and network isolation, so that the demonstrated capability cannot rely on online external assistance after challenge disclosure. The resulting evidence is interpreted relative to independently specified and calibrated matching and non-matching operating situations and is statistical rather than cryptographic.
We instantiate TP-CRIV for CNN image classifiers using probability-control-based witness generation. Experiments on ten ImageNet-pretrained TorchVision models demonstrate clear same/cross-model separation and finite-challenge verification using independently calibrated thresholds.
\end{abstract}


\section{Introduction}
Artificial intelligence (AI) models support a wide
range of applications, while their development can
require substantial computational resources, expert
knowledge, and financial investment.
Protecting the intellectual property (IP) of AI
models has therefore become an important issue
\cite{watermark-survey,Sun2023-survey}, particularly
as model theft and unauthorized reuse remain
practical concerns
\cite{tramer2016stealing,knockoff}.

At the same time, many models are deployed through Machine Learning as a Service (MLaaS), where users interact with remote APIs rather than obtaining the models themselves. In such settings, the parameters, architectures, and training records of a suspicious deployed model are generally unavailable to external parties. Consequently, even if a claimant identifies a suspicious MLaaS, an independent third party cannot directly compare the claimant's model with the deployed model to assess their relationship. Moreover, without sufficiently credible preliminary evidence supporting the claimant's allegation, the third party may have difficulty justifying a request for protocol-specific cooperation, model disclosure, or other additional assistance from the MLaaS provider. Under such circumstances, the claimant must first provide convincing technical evidence that supports the suspicion of model misappropriation and justifies further investigation.
Therefore, the fundamental research question considered in this
work is \textbf{how an independent third party can obtain
credible preliminary evidence of possible model
misappropriation without white-box access to either
model and without requiring protocol-specific
cooperation from the suspicious deployed service.}
Rather than attempting to measure global similarity
between two inaccessible models, we ask whether the
third party can issue fresh requirements whose
solutions depend strongly on model-specific behavior
of the suspicious service and evaluate whether the
claimant can satisfy those requirements using a
locally available model.

\begin{figure*}[tb]
\centering
\includegraphics[width=\linewidth]{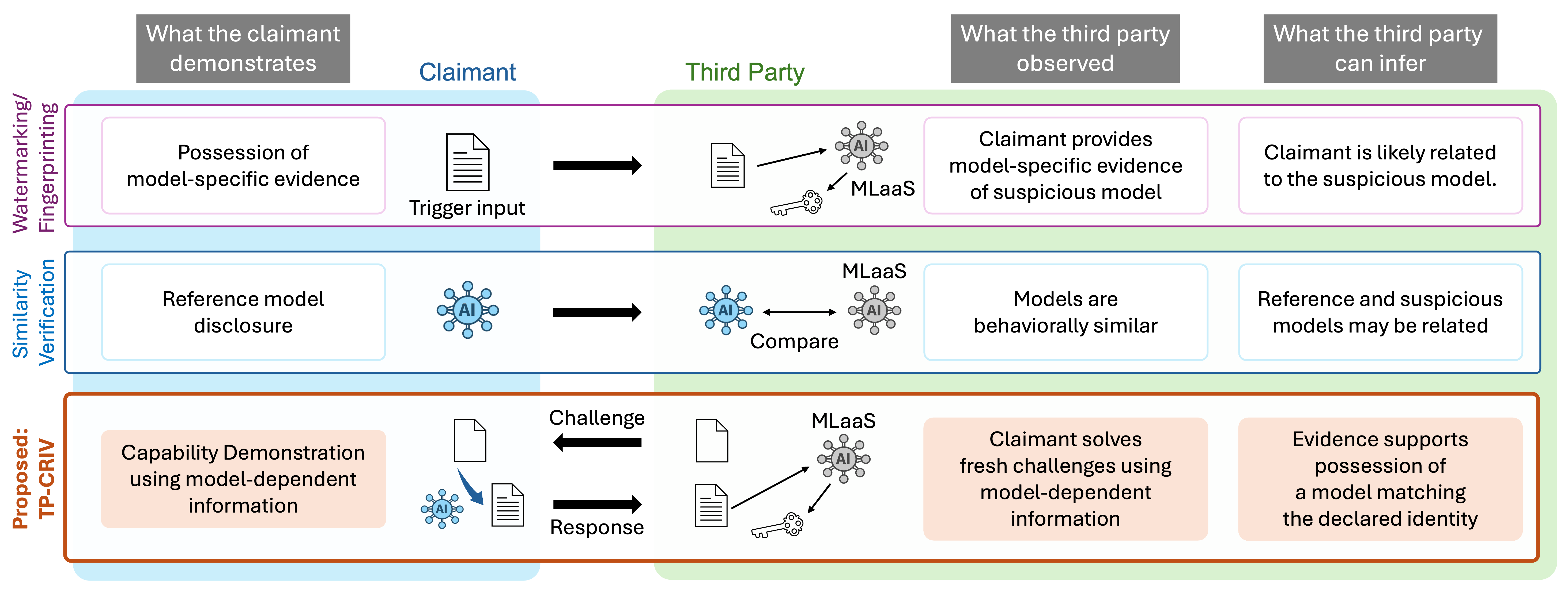}
\caption{Comparison of model-to-model verification approaches
from the perspective of the evidence available to an
independent third party and the resulting inference
about the relationship between a claimant's model
and a suspicious deployed model.}
\label{fig:comparison}
\end{figure*}

From the perspective of this research question,
existing model-protection approaches mainly use
predefined model-specific evidence, as in
watermarking and fingerprinting
\cite{watermark-survey,uchida,chen2019deepmarks,adi,rouhani2019deepsigns,frontierstitching,stablesignature,lazylayers,markyourllm,IPGuard,AFA2020,LukasZK21,DeepFoolFP,li2022dnnfp,UAPFingerprint,TRAP,QuRD},
or compare model behavior with a reference model
\cite{ModelDiff,DeepJudge,ModelLineage,ModelProvenance}.
Proof-based verifiable-inference approaches address
a different relation: rather than comparing two
models, they verify that a reported computation was
correctly performed using a specified or committed
model
\cite{SafetyNets,Mystique,zkCNN,vCNN,ZKML,zkGPT}.
These approaches provide valuable ownership,
identity, similarity, or computation-integrity
evidence, but they address a different question from
whether a claimant can use a locally available model
to satisfy fresh third-party requirements concerning
a suspicious deployed service.

To address this problem, we propose Third-Party
Challenge-Response Identity Verification (TP-CRIV)
for AI models. TP-CRIV enables an independent third party to obtain
empirical evidence as to whether a claimant possesses
a model satisfying a specified identity relative to a
suspicious deployed model, while requiring neither
white-box nor API access to the claimant's model and
only ordinary black-box access to the suspicious
MLaaS. Fig.~\ref{fig:comparison} compares this
verification objective with those of existing
model-verification approaches.
Specifically, for each intended verification problem,
the verifier first specifies two disjoint operating
situations representing matching and non-matching
configurations. These situations define the
objective-dependent model identity independently of
the challenge-response score or acceptance outcome.
The verifier then selects a model-dependent property,
issues fresh property-demanding challenges, and
evaluates the returned witnesses through the deployed
service. An operational threshold is calibrated using
independent reference configurations from the
declared matching and non-matching situations.
Freshness and network isolation exclude online
external assistance after challenge disclosure.
Within the model-based response-generation scope,
matching-consistent challenge-solving performance
provides an empirical basis for inferring that the
claimant possesses a model satisfying the specified identity. This inference is statistical rather than
cryptographic and does not establish training
provenance or legal ownership.

To clarify the role of TP-CRIV in practice, we view
model-misappropriation verification as a two-stage
process.

\textbf{Step 1: Preliminary Evidence Assessment.}
Suppose that a claimant asserts that it possesses a
model identical to the model deployed by a suspicious
MLaaS.
At this stage, however, an independent third-party
verifier may have insufficient evidence to determine
whether the claim is credible.
Under such circumstances, it may also be difficult
to justify requiring the suspicious MLaaS provider
to participate in a dedicated verification
procedure.
The verifier must therefore first assess the
technical credibility of the claimant's
misappropriation claim using externally observable
evidence and determine whether the case warrants a
more rigorous subsequent investigation.

\textbf{Step 2: Formal Ownership Investigation.}
\
If sufficient suspicion is established, a formal
investigation may employ stronger verification
conditions, including direct access to the relevant
models or protocol-specific cooperation from the
involved parties, together with ownership-related
evidence such as development records, timestamps,
and ownership documents.
Based on this technical and ownership-related evidence,
the verifier determines whether model misappropriation or
ownership infringement has occurred.

TP-CRIV addresses the first stage by providing
preliminary technical evidence about possession of a
model satisfying the declared identity.

\subsection{Our Contributions}
The main contributions of this paper are summarized
as follows:

\begin{itemize}
\item
We propose TP-CRIV for preliminary third-party
verification in suspicious MLaaS settings where the
verifier has neither white-box nor API access to the
claimant's model and the deployed service provides
only its ordinary black-box inference interface.
Under these conditions, TP-CRIV enables the verifier
to obtain empirical evidence as to whether the
claimant possesses a model satisfying the
objective-dependent identity relative to the
deployed model, without requiring protocol-specific
cooperation from the service provider.

\item
To demonstrate the practical feasibility of TP-CRIV,
we instantiate the framework for CNN-based image
classification using local input--probability
geometry and probability-control-based witness
generation
\cite{sano2025eusipco}.
Experiments with ten ImageNet-pretrained TorchVision
CNN models demonstrate clear separation between the
declared matching and non-matching configurations and
show that calibration on separate models enables
finite-challenge decisions on held-out models.
\end{itemize}

\subsection{Organization}
The remainder of this paper is organized as follows.
Section~\ref{sec:related} reviews related studies on
model protection and verification, including
ownership, identity, similarity, and proof-based
verifiable inference.
Section~\ref{sec:framework} presents the TP-CRIV
framework and its verification principle.
Section~\ref{sec:fund_eva} presents a CNN-based
instantiation and feasibility evaluation.
Section~\ref{sec:conclusion} concludes the paper and
outlines future work.

%%%%%%%%%%%%%%%%%%%%%%%%%%%

\section{Related Work}
\label{sec:related}

This section reviews existing approaches relevant to
model protection and verification, focusing on the
evidence available to an independent third party and
the conclusions that can be drawn from that evidence.

\subsection{Model-Specific Evidence for Verification}
Model watermarking and fingerprinting support ownership or identity claims using model-specific information. Although they construct such information in different ways, verification ultimately evaluates evidence associated with a particular model.

\subsubsection{Model Watermarking} Model watermarking embeds proprietary information into an AI model and verifies ownership by detecting or extracting the embedded watermark \cite{watermark-survey,uchida,chen2019deepmarks,adi,rouhani2019deepsigns,frontierstitching,stablesignature,lazylayers,markyourllm}. Representative approaches include trigger-based watermarks that induce predefined behaviors for specific inputs \cite{adi,frontierstitching,lazylayers,markyourllm} and methods that recover embedded messages from model outputs \cite{stablesignature,markyourllm}. In black-box or gray-box settings, such evidence can be verified through observable model responses without exposing the model itself. Successful verification therefore demonstrates that the presented watermark-related evidence is consistent with the suspicious model. Under the assumption that this evidence remains exclusive to the legitimate owner, such consistency can support an ownership claim. However, if the watermark information is disclosed or acquired by another party, the exclusivity assumption may be weakened. Moreover, because the watermark-related information is generally defined before verification, the verification primarily evaluates predefined evidence rather than whether the claimant currently possesses a model capable of producing fresh model-dependent responses. 

\subsubsection{Model Fingerprinting} Model fingerprinting identifies models through intrinsic model-dependent behavior without modifying the target model \cite{IPGuard,AFA2020,LukasZK21,DeepFoolFP,li2022dnnfp,UAPFingerprint,TRAP,QuRD}. Representative approaches use decision-boundary characteristics \cite{IPGuard,AFA2020,LukasZK21,DeepFoolFP}, model-specific test examples \cite{li2022dnnfp}, universal adversarial perturbations \cite{UAPFingerprint}, or model-sensitive behavior of large language models \cite{TRAP}. Such fingerprints provide evidence for determining whether the behavior of a suspicious model is consistent with model-specific information derived from a reference model. However, fingerprint evidence can generally be constructed before the verification session and may subsequently be presented or reused without requiring the claimant to demonstrate access to the model from which it was derived. Consequently, successful fingerprint verification does not by itself establish that the claimant currently possesses a model satisfying the claimed identity. 

\paragraph{Limitation of Model-Specific Evidence} Watermarking and fingerprinting provide valuable ownership and identity evidence, but both primarily verify consistency with model-specific evidence prepared before verification. Accordingly, a third party can evaluate the relationship between the presented evidence and the suspicious model, but cannot directly determine from that evidence alone whether the claimant currently possesses a model capable of generating corresponding model-dependent behavior. Furthermore, the evidence itself may be proprietary or valuable and may not always be desirable to disclose to an external verifier. These limitations motivate verification that evaluates fresh model-dependent capability rather than only possession or disclosure of predefined evidence.

\begin{table*}[t]
\centering
\renewcommand{\arraystretch}{2.0}
\caption{Comparison of verification objectives, evidence, and operational requirements.}
\label{tab:verification_objective}
\begin{tabularx}{\linewidth}{lXXXX}
\Xhline{1pt}
\textbf{Method}
& \textbf{Model-Specific Evidence}
& \textbf{Similarity Verification}
& \textbf{Proof-Based Verification}
& \textbf{Proposed: TP-CRIV}
\\
\Xhline{1pt}

\textbf{Primary objective}
&
\makecell[l]{Support ownership or\\model-identity claims}
&
\makecell[l]{Estimate or verify similarity\\between models}
&
\makecell[l]{Verify correctness of \\ a specified or \\ committed computation}
&
\makecell[l]{Third-party verification\\of objective-dependent\\identity}
\\
\hline

\makecell[l]{\textbf{Directly evaluated}\\\textbf{evidence}}
&
\makecell[l]{Consistency with predefined\\watermark or fingerprint\\evidence}
&
\makecell[l]{Behavioral or internal\\consistency between\\compared models}
&
\makecell[l]{Proof that a defined\\ computation or relation\\ is satisfied}
&
\makecell[l]{Responses to fresh\\verifier-issued requirements}
\\
\hline

\textbf{Typical approach}
&
\makecell[l]{Watermark reproduction,\\extraction, or fingerprint\\response evaluation}
&
\makecell[l]{Reference-model-based\\comparison}
&
\makecell[l]{Proof generation by the\\model-holding party}
&
\makecell[l]{Challenge-dependent\\capability demonstration}
\\
\hline

\textbf{Evidence freshness}
&
\makecell[l]{Typically predefined\\before verification}
&
\makecell[l]{Statistics obtained\\during evaluation}
&
\makecell[l]{Computation-specific proof \\generated for verification}
&
\makecell[l]{Fresh, previously\\undisclosed challenges}
\\
\hline

\makecell[l]{\textbf{Possession-related}\\\textbf{conclusion}}
&
\makecell[l]{Not explicitly\\evaluated}
&
\makecell[l]{Not explicitly\\evaluated}
&
\makecell[l]{May bind the demonstrated\\ computation \\to a committed model}
&
\makecell[l]{Empirical inference of\\possession of a matching\\model under scope}
\\
\hline

\makecell[l]{\textbf{Verifier access}\\\textbf{to claimant model}}
&
Not required
&
\makecell[l]{Often required through\\API or white-box access}
&
Not required
&
Not required
\\
\hline

\makecell[l]{\textbf{Dedicated participation}\\
\textbf{by deployed service}}
&
\makecell[l]{Not generally required\\beyond ordinary responses}
&
\makecell[l]{Not generally required,\\but comparison access\\ is needed}
&
\makecell[l]{Typically required from\\ the model-holding service}
&
\makecell[l]{Not required beyond ordinary\\black-box inference}
\\

\Xhline{1pt}
\end{tabularx}
\end{table*}

\subsection{Model Similarity Verification}
Several studies verify model similarity or provenance
by comparing observable or internal model behavior
\cite{ModelDiff,DeepJudge,ModelLineage,ModelProvenance}.
Representative approaches include test-case-based
comparison \cite{ModelDiff,DeepJudge}, model lineage
analysis using internal information
\cite{ModelLineage}, and provenance testing based on
observable outputs \cite{ModelProvenance}.
These methods can provide strong evidence about the
relationship between models, but generally assume
that the models to be compared are available to the
verifier at least through their observable behavior,
and some methods use white-box reference-model
information to construct model-sensitive test cases
\cite{DeepJudge}.
In practical settings, however, the claimant may
also be unable to provide its model to a third party,
for example because the model is proprietary or too
large to make such access practical.
Under such conditions, direct model-similarity
verification becomes difficult.

TP-CRIV targets a different operational setting in
which the claimant's model is not made available to
the verifier, even through query access, and the
suspicious deployed service is accessible only
through its ordinary black-box interface.

\subsection{Proof-based Verifiable Inference}

Proof-based verifiable-inference methods address a
different verification objective from model-to-model
identity verification. Their primary goal is to prove
that a reported output was correctly computed from a
given input using a specified or committed model,
without necessarily revealing the model parameters
\cite{SafetyNets,Mystique,zkCNN,vCNN,ZKML,zkGPT}.
Thus, these methods primarily establish a
cryptographic relation between a model and its
computation, rather than directly evaluating the
relationship between a claimant-held model and a
separately deployed suspicious model.

In typical verifiable-inference settings, the
model-holding service acts as the prover and
participates in a dedicated verification procedure,
for example by committing to the model and generating
an inference proof.
Such mechanisms can provide strong guarantees when
this participation is available.
However, as discussed above, they become difficult
to apply when requiring the suspicious MLaaS
provider to participate in a dedicated verification
procedure is impractical.
TP-CRIV instead evaluates the relationship between
two separately held models through only ordinary
black-box access to the suspicious MLaaS, enabling
the verifier to infer the claimant's condition and
assess whether the claimant's identity claim is
technically supported.

%%%%%%%%%%%%%%%%%%%%%%%%%%%
%%%%%%%%%%%%%%%%%%%%%%%%%%%

\section{Proposed Framework}
\label{sec:framework}

\subsection{Motivation and Positioning}
TP-CRIV is designed to provide an independent third party with empirical evidence about the relationship between a claimant's locally possessed model and a suspicious remotely deployed model when direct model comparison is unavailable. Through fresh challenge-response interactions, the framework enables the verifier to assess whether the claimant demonstrates behavior consistent with the identity specified for the intended verification objective, without requiring access to the claimant's model or protocol-specific cooperation from the deployed service. To make this inference well defined, TP-CRIV separates four concepts that must not be
conflated.
First, the verification objective specifies an
a priori distinction between matching and
non-matching operating situations.
Second, a model-dependent property supplies
information for constructing discriminative
requirements but does not define model identity.
Third, fresh property-demanding challenges elicit
observable challenge-solving capability rather than
directly revealing or comparing the property.
Fourth, calibration on independent reference
configurations provides an operational interpretation
of the resulting scores relative to the predeclared
identity distinction.

These roles are formalized below.
Compared with model-specific evidence approaches,
TP-CRIV evaluates not only evidence associated with
the claimant and the suspicious service, but also
whether the claimant can use a locally available
model to satisfy fresh model-dependent requirements.
In contrast to model-similarity verification,
TP-CRIV does not require the verifier to access or
query the claimant's model.
Unlike proof-based verifiable inference, it also
does not require the suspicious MLaaS provider to
participate in a dedicated verification procedure.
Table~\ref{tab:verification_objective} summarizes
these differences from existing verification
approaches.

\subsection{System Model}

\begin{figure*}[tb]
\centering
\includegraphics[width=\textwidth]{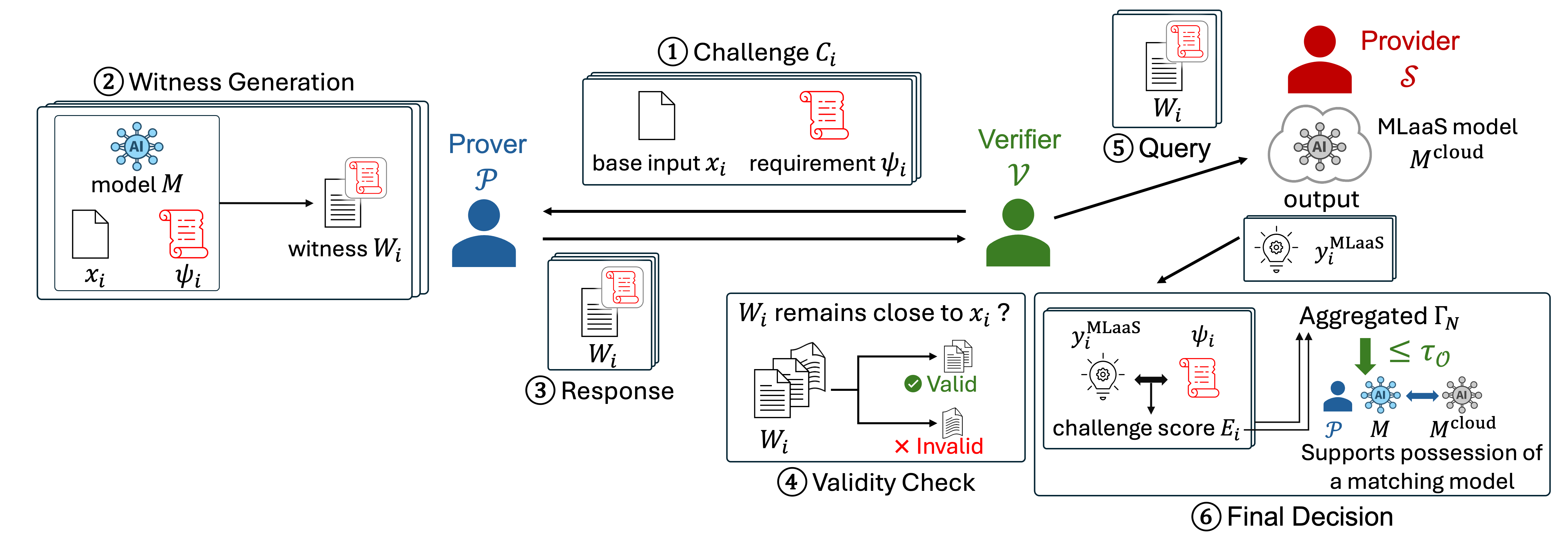}
\caption{
Overview of the TP-CRIV framework.
The verifier specifies the objective-dependent
model-identity distinction, selects a model-dependent
property, issues fresh property-demanding challenges,
and evaluates the prover's returned responses through
the deployed MLaaS.
The prover answers the challenges using a locally
available candidate model.
}
\label{fig:prop_frame}
\end{figure*}

We consider three entities:
\begin{itemize}
\item
\textbf{Prover} $\mathcal{P}$,
which locally possesses a candidate model $M$ and
claims that $M$ corresponds to the matching situation
relative to $M^{\mathrm{cloud}}$.
\item
\textbf{Service provider} $\mathcal{S}$,
which operates $M^{\mathrm{cloud}}$ through an MLaaS
platform.
\item
\textbf{Verifier} $\mathcal{V}$,
which specifies a verification objective
$\mathcal{O}$, prepares or obtains calibration data,
conducts verification, and evaluates the resulting
evidence.
\end{itemize}
Let $F^{\mathrm{MLaaS}}$ denote the black-box MLaaS
interface exposed by $\mathcal{S}$.
For an input $x$, $\mathcal{V}$ observes only
\begin{equation}
y^{\mathrm{MLaaS}}
=
F^{\mathrm{MLaaS}}(x),
\end{equation}
where, in the baseline framework setting,
\begin{equation}
F^{\mathrm{MLaaS}}(x)
=
M^{\mathrm{cloud}}(x).
\end{equation}
Neither $\mathcal{P}$ nor $\mathcal{V}$ directly
accesses $M^{\mathrm{cloud}}$.
Moreover, $\mathcal{V}$ has no direct access to $M$
and need not observe its parameters, architecture,
or other internal information.
Consequently, $\mathcal{V}$ cannot directly compare
the internal structures of $M$ and
$M^{\mathrm{cloud}}$.
TP-CRIV therefore evaluates the claimed relationship
from observable interactions with the deployed
MLaaS under the declared verification objective,
calibration procedure, and operational assumptions.

\subsection{Verification Scope and Operational Assumptions}

\paragraph{Non-Matching Model Configurations}

A model configuration declared non-matching for
$\mathcal{O}$ constitutes the negative identity case.
Depending on the intended objective, $M$ may be
unrelated to, independently trained from, extracted
from \cite{tramer2016stealing}, distilled from \cite{Hinton2015Distillation}, or otherwise derived from
$M^{\mathrm{cloud}}$.
Such provenance does not determine the identity
label; the label is fixed by the operating situations
declared for $\mathcal{O}$.
Acceptance of a declared non-matching configuration
constitutes a false acceptance.

Model configurations outside the declared operating
population are outside the characterized identity
scope.
Accordingly, the reported verification performance
does not automatically extend to model configurations
that are not represented in the corresponding
calibration, evaluation, or supporting analysis.

\paragraph{Service-Side Assumption}

$\mathcal{S}$ does not disclose
$M^{\mathrm{cloud}}$, its parameters, architecture,
or inference pipeline.
$\mathcal{V}$ interacts with
$F^{\mathrm{MLaaS}}$ only through its ordinary
black-box interface.
$\mathcal{S}$ is assumed to return the ordinary output
of a fixed deployed inference pipeline for each
submitted input.
Adaptive routing of verification queries, deliberate
output manipulation, and query refusal are outside
the present scope.
The verification decision therefore concerns the
observed behavior of the deployed service and does
not establish its internal implementation.

\paragraph{Network Isolation}

Before the first challenge of a verification session
is disclosed, $\mathcal{P}$ is placed under network
isolation.
The isolation remains in effect until all responses
for the session have been returned.
During this interval, $\mathcal{P}$ may use the
disclosed challenges, its locally available model
$M$, and local computational resources, but may not
obtain additional challenge-relevant information
from $F^{\mathrm{MLaaS}}$, another remote model, a
relay service, an external oracle, or any other
external source.
Thus, network isolation excludes online external
assistance after challenge disclosure.

\subsection{TP-CRIV Protocol}

A verification trial consists of $N$
challenge-response interactions.
For each interaction $i$, $\mathcal{V}$ issues a
fresh challenge, $\mathcal{P}$ generates a
corresponding response using its locally available
model, and $\mathcal{V}$ evaluates the response
through the deployed MLaaS.

Let $\operatorname{ResGen}$ denote the response-generation algorithm
used by $\mathcal{P}$.
The present framework adopts a model-based
response-generation scope:
$\operatorname{ResGen}$ may use $M$, the disclosed challenge, and local
computation required to operate or process $M$.
A standalone learned or precomputed mechanism that
can generate responses without using $M$ is outside
the present model-based response-generation scope.

Fig.~\ref{fig:prop_frame} illustrates the overall TP-CRIV framework.
The $i$-th interaction proceeds as follows.

\begin{enumerate}
\item
$\mathcal{V}$ generates a fresh challenge
\begin{equation}
C_i=(x_i,\psi_i),
\end{equation}
where $x_i$ is a base input and $\psi_i$ specifies
the verification requirement.

\item
Given $C_i$, $\mathcal{P}$ applies $\operatorname{ResGen}$ to its local
model $M$ and generates a witness
\begin{equation}
W_i\leftarrow \operatorname{ResGen}(M,C_i).
\label{eq:model_response_generation}
\end{equation}
\item
$\mathcal{P}$ then returns $W_i$ to
$\mathcal{V}$.

\item
$\mathcal{V}$ evaluates the challenge-specific
witness validity
\begin{equation}
\operatorname{Valid}_{\mathrm{wit}}(C_i,W_i)
\in\{0,1\}.
\end{equation}
An invalid witness is excluded from subsequent score
evaluation.

\item
For each valid $W_i$, $\mathcal{V}$ submits $W_i$
to the MLaaS and obtains
\begin{equation}
y_i^{\mathrm{MLaaS}}
=
F^{\mathrm{MLaaS}}(W_i).
\end{equation}

\item
After the $N$ interactions, $\mathcal{V}$ defines
\begin{equation}
\mathcal{I}_{\mathrm{val}}
=
\left\{
i\mid
\operatorname{Valid}_{\mathrm{wit}}(C_i,W_i)=1
\right\},
\end{equation}
and
\begin{equation}
N_{\mathrm{val}}
=
\left|\mathcal{I}_{\mathrm{val}}\right|.
\end{equation}

Let $N_{\min}$ denote the minimum number of valid
witnesses required for score aggregation.
If $N_{\mathrm{val}}<N_{\min}$, the session is
rejected.
Otherwise, for each
$i\in\mathcal{I}_{\mathrm{val}}$,
$\mathcal{V}$ computes
\begin{equation}
E_i
=
\operatorname{Eval}(C_i,W_i,y_i^{\mathrm{MLaaS}}),
\end{equation}
where $\operatorname{Eval}$ denotes the challenge-specific
scoring function, and $E_i$ is the resulting
per-challenge score.

Let $\operatorname{Aggr}$ denote the declared aggregation rule.
The valid scores are aggregated as
\begin{equation}
\Gamma_N
=
\operatorname{Aggr}\left(
\{E_i\}_{i\in\mathcal{I}_{\mathrm{val}}}
\right).
\label{eq:general_aggregate_score}
\end{equation}

Let $\tau_{\mathcal{O}}$ denote the operational
acceptance threshold fixed for $\mathcal{O}$ before
the suspicious configuration is evaluated.
The final decision is
\begin{equation}
\mathsf{Accept}
=
\begin{cases}
0,
& N_{\mathrm{val}}<N_{\min},
\\
\mathbb{I}\!\left[
\Gamma_N\leq\tau_{\mathcal{O}}
\right],
& N_{\mathrm{val}}\geq N_{\min}.
\end{cases}
\label{eq:general_acceptance_rule}
\end{equation}
\end{enumerate}
$\tau_{\mathcal{O}}$ is calibrated using data
prepared independently of the suspicious
configuration.
Here, $\mathbb{I}[\cdot]$ denotes the indicator function,
which returns 1 if the enclosed condition is true and 0 otherwise.
Its reported operating characteristics apply only
to the model population, challenge distribution,
response-generation conditions, and deployment
conditions represented in the supporting calibration
and evaluation.

In particular, different choices of $\operatorname{ResGen}$ may induce
different score distributions for the same
$(M,M^{\mathrm{cloud}})$ configuration.
Accordingly, a reported soundness or
false-acceptance characterization covers only the
choices of $\operatorname{ResGen}$ represented in the corresponding
evaluation or supported by separate analysis.
Standalone non-model response mechanisms remain
outside the present model-based soundness scope.

The interpretation of acceptance under freshness,
network isolation, and the stated model-based scope
is developed in
Section~\ref{section:verification_principle}.
\subsection{Verification Principle}
\label{section:verification_principle}

TP-CRIV separates four logically distinct elements:
the objective-dependent identity distinction, the
model-dependent property used to construct
challenges, the operational characterization of
challenge-solving performance, and the inference
obtained by combining matching consistency with the
freshness and network-isolation conditions.
This separation prevents the verification outcome
from defining the identity relation that it is
intended to test.

\subsubsection{Operating Situations and Objective-Dependent Identity}
For each intended verification problem,
$\mathcal{V}$ specifies two disjoint operating
situations,
$\Pi_{1}^{\mathcal{O}}$ and
$\Pi_{0}^{\mathcal{O}}$,
representing configurations regarded as matching and
non-matching, respectively.
These situations are specified a priori
according to the semantic distinction relevant to
the intended application and independently of the
challenge-response score, calibration threshold, or
acceptance outcome.
Together, they define the model-identity distinction
tested by the verification objective $\mathcal{O}$.

The corresponding operating population is
\begin{equation}
\Pi^{\mathcal{O}}
=
\Pi_{1}^{\mathcal{O}}\cup\Pi_{0}^{\mathcal{O}},
\qquad
\Pi_{1}^{\mathcal{O}}\cap\Pi_{0}^{\mathcal{O}}=\emptyset.
\label{eq:operating_population_partition}
\end{equation}
Thus, model identity is objective dependent and
population scoped.
TP-CRIV does not posit a universal identity relation
over all possible AI models.

For convenience, membership in these two situations
is encoded by
\begin{equation}
\operatorname{Match}_{\mathcal{O}}\left(M,M^{\mathrm{cloud}}\right)
=
\begin{cases}
1,&(M,M^{\mathrm{cloud}})\in\Pi_{1}^{\mathcal{O}},\\
0,&(M,M^{\mathrm{cloud}})\in\Pi_{0}^{\mathcal{O}}.
\end{cases}
\label{eq:objective_match_predicate}
\end{equation}
Here, $\operatorname{Match}_{\mathcal{O}}$ introduces
no additional notion of identity and is not inferred
from the TP-CRIV score.
It only records membership in the two situations
specified independently for $\mathcal{O}$.
For example, one verification objective may declare
same-model-instance configurations as matching and
selected distinct model instances as non-matching.
Within the corresponding $\Pi^{\mathcal{O}}$, this
constitutes an exact-instance distinction.
Another objective may place designated descendant
models in the matching situation and selected
outside-lineage models in the non-matching situation,
yielding a lineage-oriented distinction.
The meaning of identity therefore changes when the
declared operating situations change.

A configuration outside $\Pi^{\mathcal{O}}$ is
outside the identity distinction instantiated by
$\mathcal{O}$.
The protocol may still produce an observable score
for such a configuration, but the current objective
does not assign it a matching or non-matching label,
and the reported operating characteristics do not
automatically apply to it.
Extending the identity claim to such a configuration
requires extending the declared operating situations
and establishing corresponding calibration,
evaluation, or analysis.

The choice of $\operatorname{ResGen}$ does not define model identity.
The matching and non-matching labels are fixed
a priori by $\Pi_{1}^{\mathcal{O}}$ and
$\Pi_{0}^{\mathcal{O}}$.
Different choices of $\operatorname{ResGen}$ may affect observable
challenge-solving performance and operating error
rates, but they do not change those predeclared
identity labels.

\subsubsection{Model-Dependent Property and Challenge-Solving Capability}
For each model $M$, let $S_M$ denote a selected
model-dependent property.
Examples include local decision geometry in a
classifier, conditional generation behavior in a
generative model, or another high-dimensional
behavioral structure whose detailed form depends on
the model.
The selected property does not define
$\Pi_{1}^{\mathcal{O}}$ or
$\Pi_{0}^{\mathcal{O}}$.
Instead, $\mathcal{V}$ uses it as a source of
model-dependent information from which fresh
verification requirements can be constructed.

For $\Pi^{\mathcal{O}}$, a useful property is one
whose challenge-relevant information enables
configurations in $\Pi_{1}^{\mathcal{O}}$ to solve
requirements determined by $M^{\mathrm{cloud}}$ more
consistently than configurations in
$\Pi_{0}^{\mathcal{O}}$ under the covered choices
of $\operatorname{ResGen}$.
For a challenge $C=(x,\psi)$, the requirement $\psi$
is constructed so that a valid low-error solution
depends substantially on challenge-relevant aspects
of $S_{M^{\mathrm{cloud}}}$.
Such a challenge is referred to as a
property-demanding challenge.

For each valid witness $W$, $\mathcal{V}$ evaluates
the observable score
$\operatorname{Eval}\left(C,W,F^{\mathrm{MLaaS}}(W)\right),$
which measures how well the returned witness
satisfies the issued requirement.
$\mathcal{V}$ need not reconstruct $S_M$,
$S_{M^{\mathrm{cloud}}}$, or the complete set of
successful witnesses.
It observes only whether $\mathcal{P}$ can use $M$
through $\operatorname{ResGen}$ to generate successful responses to
newly issued challenges.
This capability relation is not deterministic.
A configuration in $\Pi_{1}^{\mathcal{O}}$ may
occasionally fail because of numerical, algorithmic,
or deployment-side effects.
Conversely, a configuration in
$\Pi_{0}^{\mathcal{O}}$ may reproduce sufficient
challenge-relevant information to solve some or all
challenges.
Such success does not change its predeclared class
membership.
If it is accepted, the event constitutes a false
acceptance under the declared objective.

Accordingly, the selected property provides the
mechanism through which model-dependent information
is elicited as an observable challenge-solving
capability.
Whether this capability actually discriminates the
two declared situations must be established
operationally rather than assumed from a converse
relation between property similarity and model
identity.

\subsubsection{Operational Characterization of Challenge-Solving Performance}
Let $\mathcal{D}_C$ denote the declared distribution
from which fresh challenges are drawn, and let $\operatorname{ResGen}$
denote the choice covered by the operating
characterization.
For each $b\in\{0,1\}$, let
$Q_b^{\mathcal{O}}$ denote the declared distribution
over model configurations in
$\Pi_b^{\mathcal{O}}$ used for operational
characterization.
Sampling
\begin{equation}
(M,M^{\mathrm{cloud}})
\sim
Q_b^{\mathcal{O}},
\end{equation}
drawing fresh challenges from $\mathcal{D}_C$,
generating responses using $\operatorname{ResGen}$, and applying
Eq.~\eqref{eq:general_aggregate_score}
induce the observable aggregate-score distribution
\begin{equation}
\Gamma_N
\sim
P_{b,\operatorname{ResGen}}^{\mathcal{O}}.
\label{eq:conditional_score_distribution}
\end{equation}
These distributions provide an operational
characterization of the challenge-solving capability
demonstrated by the matching and non-matching
populations under the specified conditions.
The purpose of the selected property and challenge
family is therefore to induce sufficiently
distinguishable observable performance between
$\Pi_{1}^{\mathcal{O}}$ and
$\Pi_{0}^{\mathcal{O}}$.

TP-CRIV does not assume a deterministic, one-to-one,
or monotone mapping between a latent discrepancy
between $S_M$ and $S_{M^{\mathrm{cloud}}}$ and the
observed aggregate score $\Gamma_N$.
Different model configurations may produce similar
scores, the same configuration may produce different
scores for different sets of fresh challenges, and
different choices of $\operatorname{ResGen}$ may induce different score
distributions for the same
model configuration.
The relevant question is instead whether the
challenge-solving performance induced by the
declared matching and non-matching populations is
sufficiently separated for the intended operating
criterion.

Using independent calibration data, $\mathcal{V}$
determines $\tau_{\mathcal{O}}$ before evaluating the
suspicious configuration.
Let $RG_0$ and $RG_1$ denote the covered choices of
$\operatorname{ResGen}$ for the non-matching and matching situations,
respectively; they may coincide.
The corresponding false-acceptance rate ($\operatorname{FAR}_{\mathcal{O},RG_0}$) and
false-rejection rate ($\operatorname{FRR}_{\mathcal{O},RG_1}$) are given as follows:
\begin{equation}
\operatorname{FAR}_{\mathcal{O},RG_0}
=
\Pr_{\substack{
(M,M^{\mathrm{cloud}})
\sim Q_{0}^{\mathcal{O}}
\\
C_{1:N}\sim\mathcal{D}_C^{N}
}}
\left[
\Gamma_N\leq\tau_{\mathcal{O}}
\mid RG_0
\right],
\label{eq:operational_far}
\end{equation}
and
\begin{equation}
\operatorname{FRR}_{\mathcal{O},RG_1}
=
\Pr_{\substack{
(M,M^{\mathrm{cloud}})
\sim Q_{1}^{\mathcal{O}}
\\
C_{1:N}\sim\mathcal{D}_C^{N}
}}
\left[
\Gamma_N>\tau_{\mathcal{O}}
\mid RG_1
\right].
\label{eq:operational_frr}
\end{equation}
The probabilities include the randomness of model
configuration sampling, fresh challenge generation,
and, when applicable, the response-generation
algorithm.
$\tau_{\mathcal{O}}$ provides an operational
boundary for interpreting observed challenge-solving
performance relative to the declared
$\Pi_{1}^{\mathcal{O}}$ and
$\Pi_{0}^{\mathcal{O}}$ situations.

Within the declared operating population,
acceptance provides finite-sample empirical evidence
that the observed challenge-solving performance is
more consistent with the characterized matching
population than with the characterized non-matching
population.
TP-CRIV evaluates evidence for membership
in a predeclared identity class rather than defining
identity from acceptance.

The reported $\tau_{\mathcal{O}}$, FAR, and FRR
apply only to the declared operating distributions,
$\mathcal{D}_C$, $\operatorname{ResGen}$, and deployment conditions.
Broader claims require corresponding extension of
the operating characterization.

\subsubsection{Inference Under Freshness and Network Isolation}

The calibrated decision characterizes whether the
observed challenge-solving performance is consistent
with the declared matching situation; by itself, it
does not establish whether the response was generated
without external assistance.
Fresh challenges limit replay based on a small
precomputed response set, while the network-isolation
condition prevents $\mathcal{P}$ from obtaining
additional challenge-relevant information from
external sources after disclosure.

Within the stated model-based response-generation
scope, the session responses are generated through
$\operatorname{ResGen}(M,C)$.
If $\Gamma_N$ also lies in the calibrated matching
region, the observed capability is characteristic of
the matching situation within the declared operating
scope.
These conditions jointly provide an empirical basis
for inferring that $\mathcal{P}$ possesses a model
satisfying the identity specified by $\mathcal{O}$.
The inference is conditional on the stated scope and
does not establish cryptographic binding, training
provenance, or legal ownership.

The verifier's inference process is summarized in
Table~\ref{tab:verification_inference}.

\begin{table}[t]
\centering
\caption{Verifier's inference process in TP-CRIV.}
\label{tab:verification_inference}
\renewcommand{\arraystretch}{1.35}
\begin{tabularx}{\columnwidth}{@{}cX@{}}
\toprule
\textbf{Step}
&
\textbf{Verifier's observation or inference}
\\
\midrule

1
&
$\mathcal{P}$ repeatedly returns valid low-score
witnesses for fresh property-demanding challenges.
\\

$\Downarrow$
&
\textit{The transcript directly demonstrates finite challenge-solving capability.}
\\

2
&
Freshness limits replay-based strategies, while
network isolation excludes online external
assistance after challenge disclosure.
\\

$\Downarrow$
&
\textit{$\mathcal{P}$ locally possesses the
challenge-solving capability required to generate
the demonstrated responses without online external
assistance.}
\\

3
&
Under the model-based response-generation scope,
the aggregate score lies in the acceptance region
calibrated for the declared matching situation.
\\

$\Downarrow$
&
\textit{The candidate model exhibits
matching-consistent challenge-solving performance
within the calibrated scope.}
\\

4
&
The combined evidence supports an empirical
inference that the claimant possesses a model
satisfying the objective-dependent identity.
\\

\bottomrule
\end{tabularx}
\end{table}

\subsection{Protocol Requirements}

A valid TP-CRIV instantiation should satisfy the
following requirements under
Eq.~\eqref{eq:general_acceptance_rule}.

\paragraph{Objective and Population Specification}
$\mathcal{V}$ must specify the disjoint situations
$\Pi_{1}^{\mathcal{O}}$ and
$\Pi_{0}^{\mathcal{O}}$ before evaluating the
suspicious case, independently of the TP-CRIV score,
$\tau_{\mathcal{O}}$, or the eventual decision.
$\mathcal{V}$ must also specify the operating
distributions $Q_{1}^{\mathcal{O}}$ and
$Q_{0}^{\mathcal{O}}$, $\mathcal{D}_C$, the choices
of $\operatorname{ResGen}$ covered by calibration and evaluation, and
the criterion used to select $\tau_{\mathcal{O}}$.
$\Pi_{0}^{\mathcal{O}}$ should contain the model
configurations that constitute relevant alternatives
to the intended identity claim; omitted model classes
and unrepresented algorithms remain outside the
characterized soundness scope.

\paragraph{Completeness}
Let $\epsilon_c$ denote the allowed completeness
error.
For $(M,M^{\mathrm{cloud}})\sim
Q_{1}^{\mathcal{O}}$ and a covered $\operatorname{ResGen}$, acceptance
should occur with high probability:
\begin{equation}
\Pr_{\substack{
(M,M^{\mathrm{cloud}})\sim Q_{1}^{\mathcal{O}}
\\
C_{1:N}\sim\mathcal{D}_C^{N}
}}
\left[\mathsf{Accept}=1\mid \operatorname{ResGen}\right]
\geq 1-\epsilon_c.
\label{eq:completeness}
\end{equation}
Here, $\epsilon_c$ is the objective-specific
false-rejection probability under the declared
operating conditions.

\paragraph{Soundness}
Let $\epsilon_s$ denote the allowed soundness error.
For $(M,M^{\mathrm{cloud}})\sim
Q_{0}^{\mathcal{O}}$ and a covered $\operatorname{ResGen}$, acceptance
should occur with low probability:
\begin{equation}
\Pr_{\substack{
(M,M^{\mathrm{cloud}})\sim Q_{0}^{\mathcal{O}}
\\
C_{1:N}\sim\mathcal{D}_C^{N}
}}
\left[\mathsf{Accept}=1\mid \operatorname{ResGen}\right]
\leq \epsilon_s.
\label{eq:soundness}
\end{equation}
Here, $\epsilon_s$ is the objective-specific
false-acceptance probability under the declared
operating conditions.
The identity label is independent of $\operatorname{ResGen}$; therefore,
acceptance of any configuration assigned to
$\Pi_{0}^{\mathcal{O}}$ is a false acceptance.

\paragraph{Property-Demanding Challenge Design}
Each challenge
$C=(x,\psi)$
must contain a fresh requirement whose successful
solution depends substantially on model-dependent
information associated with
$S_{M^{\mathrm{cloud}}}$.
The challenge family should separate the induced
score distributions of the relevant matching and
non-matching populations at the intended operating
point.
A fixed witness, a small precomputed response set, or
challenge-independent evidence should not yield
acceptance with high probability over fresh
challenges.

\paragraph{Calibration and Threshold Transfer}
$\tau_{\mathcal{O}}$ must be determined from
calibration data prepared independently of the
suspicious configuration and fixed before that
configuration is evaluated.
Reported error rates are valid only under the
challenge-generation process, validity conditions,
aggregation rule, $\operatorname{ResGen}$, and deployment pipeline
represented in calibration and evaluation.

\paragraph{Freshness and Network Isolation}
Each $C_i\sim\mathcal{D}_C$ must remain
unpredictable before disclosure, and the challenge
family should limit reuse of responses across fresh
challenges.
The network-isolation condition defined above must
hold throughout witness generation.
Both conditions are required for the inference of
Section~\ref{section:verification_principle}.

\section{CNN-Based Instantiation and Feasibility Evaluation}
\label{sec:fund_eva}

This section instantiates the abstract TP-CRIV
components for CNN-based image classification under
an explicitly bounded experimental setting.
The evaluation fixes
$\operatorname{ResGen}=\mathrm{I\mathchar`-FMPC}$ for all tested
$(M,M^{\mathrm{cloud}})$ configurations and examines
whether the declared matching and non-matching
situations induce separable scores and whether a
threshold calibrated on separate models transfers to
held-out configurations.
The results characterize only the operating scope
instantiated below.

\subsection{Verification Objective and Operating Scope}

We consider a CNN-based feasibility demonstration of
TP-CRIV under an explicitly bounded operating scope.
Both $M$ and $M^{\mathrm{cloud}}$ are CNN image
classifiers.
Let $\mathcal{M}_{\mathrm{eval}}$ denote the finite
model pool consisting of ten ImageNet-pretrained CNN
model instances:
ResNet-18, -34, -101, and -152;
VGG-11, -13, -16, and -19;
and two ResNet-50 instances using the TorchVision
\text{IMAGENET1K\_V1} and
\text{IMAGENET1K\_V2} pretrained weights,
respectively
\cite{ResNet50,VGG,PyTorch}.
The two ResNet-50 instances are hereafter denoted by
ResNet-50 $(\alpha)$ and ResNet-50 $(\beta)$.
They share the same architecture but contain
different pretrained parameter sets obtained using
different training recipes.
The evaluated pool is a heterogeneous set of
publicly available pretrained model instances that
differ primarily in architecture family, network
depth, or pretrained weight configuration.
All models perform the same ImageNet classification
task and share the same output-label space.

For the present feasibility study, the operating
population is declared as
\begin{equation}
\Pi^{\mathcal{O}}
=
\mathcal{M}_{\mathrm{eval}}\times\mathcal{M}_{\mathrm{eval}}.
\label{eq:cnn_operating_population}
\end{equation}
Thus, $\Pi^{\mathcal{O}}$ contains all directed
$(M,M^{\mathrm{cloud}})$ configurations formed from
the ten models in $\mathcal{M}_{\mathrm{eval}}$.
Before challenge-response evaluation, the
verification objective $\mathcal{O}$ declares the
following two situations within
$\Pi^{\mathcal{O}}$.
These declarations are independent of the observed
scores and thresholds.

\paragraph{Matching situation}
The first situation is
\begin{equation}
\Pi_{1}^{\mathcal{O}}
=
\left\{(M,M^{\mathrm{cloud}})\in\Pi^{\mathcal{O}}
\;\middle|\;M=M^{\mathrm{cloud}}\right\}.
\label{eq:cnn_matching_population}
\end{equation}
It contains the 10 same-model directed
configurations, which are designated as the
identity-satisfying, or matching, situation for the
present objective.

\paragraph{Non-matching situation}
The contrasting situation is
\begin{equation}
\Pi_{0}^{\mathcal{O}}
=
\left\{(M,M^{\mathrm{cloud}})\in\Pi^{\mathcal{O}}
\;\middle|\;M\neq M^{\mathrm{cloud}}\right\}.
\label{eq:cnn_nonmatching_population}
\end{equation}
It contains the 90 cross-model directed
configurations formed from distinct model instances
in $\mathcal{M}_{\mathrm{eval}}$, which are
designated as the non-matching situation.
For the finite operating characterization below,
$Q_{1}^{\mathcal{O}}$ and $Q_{0}^{\mathcal{O}}$ are
taken as uniform distributions over the 10 and 90
configurations, respectively.

In the present feasibility study, both matching and non-matching provers use $\operatorname{ResGen}=\mathrm{I\mbox{-}FMPC}$. Thus, the reported score distributions and FAR characterize non-matching models that attempt the same fine-grained probability-control task as matching models. Other model-based choices of $\operatorname{ResGen}$ are outside the present operating characterization.

Together, these situations define an exact-instance
distinction scoped to $\Pi^{\mathcal{O}}$.
Models outside $\mathcal{M}_{\mathrm{eval}}$,
including independently trained, fine-tuned \cite{Yosinski2014Transfer}, pruned \cite{Han2015Pruning},
quantized \cite{Jacob2018Quantization}, extracted, or distilled models, are not
assigned a label by this experimental objective and
are not characterized by the reported FAR.
Each evaluated $M$ is used directly with the fixed
$\operatorname{ResGen}$ specified above.

\subsection{Verification Procedure}

\begin{figure*}[tb]
\centering
\includegraphics[width=\textwidth]{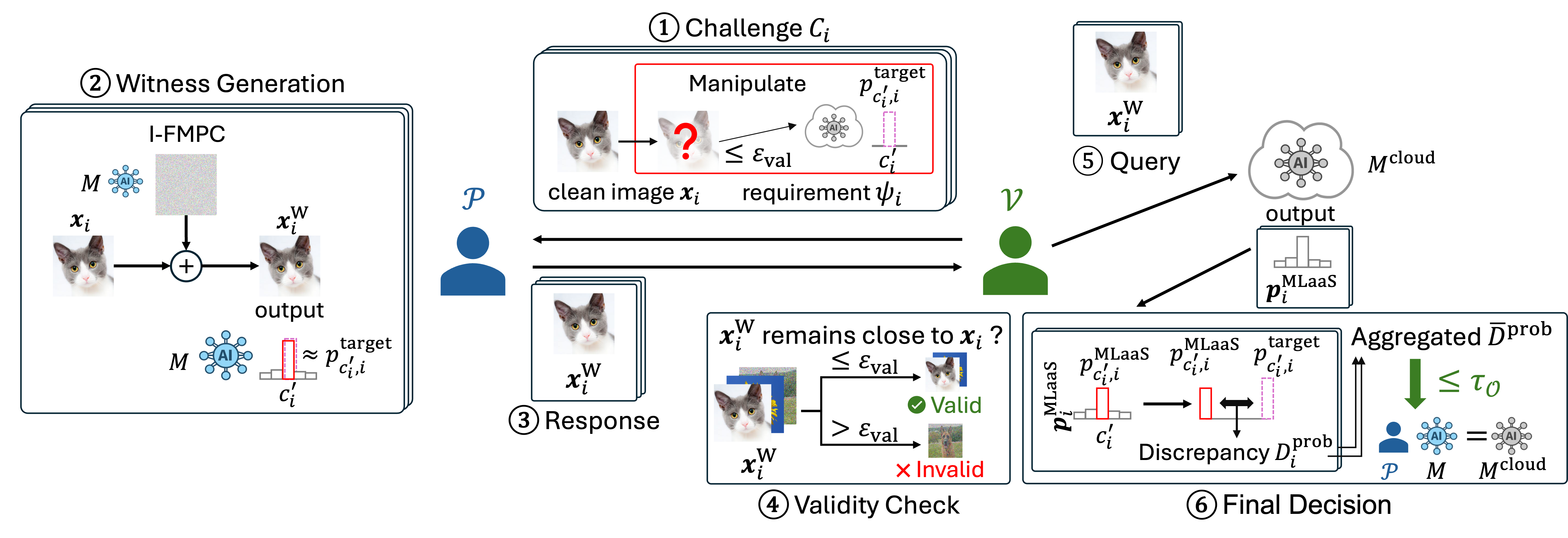}
\caption{
Verification procedure in the image-classification
instantiation of TP-CRIV.
$\mathcal{V}$ specifies an input image, a fresh
probability-control requirement, and a
witness-modification constraint.
}
\label{fig:cnn_verification_procedure}
\end{figure*}

Fig.~\ref{fig:cnn_verification_procedure}
illustrates the verification procedure.
The selected property $S_M$ is the local
input--probability geometry induced by CNN model
$M$, namely the local relationship between bounded
input modifications and the resulting class
probability changes.
This property guides the construction of challenges
whose valid low-error solutions depend on
$S_{M^{\mathrm{cloud}}}$.

For each verification interaction,
$\mathcal{V}$ generates a challenge
\begin{equation}
C=(\bvec{x},\psi),
\end{equation}
where $\bvec{x}$ is a clean input image and
\begin{equation}
\psi
=
\left(
c',
p^{\mathrm{target}}_{c'},
\varepsilon_{\mathrm{val}}
\right).
\label{eq:cnn_challenge_requirement}
\end{equation}
Let $\mathcal{Y}$ denote the output-label space.
Here, $\mathcal{V}$ selects the target class $c'\in\mathcal{Y}$ and specifies the required output probability $p^{\mathrm{target}}_{c'}$ and the maximum permitted witness modification $\varepsilon_{\mathrm{val}}$.
The $\mathcal{D}_C$ used in
the experiments is instantiated as follows.
$\bvec{x}$ is randomly selected from
the ImageNet dataset \cite{ImageNet}, $c'$ is randomly selected from the full set of
1000 ImageNet classes, and
$p^{\mathrm{target}}_{c'}$ is randomly selected from
$\{0.10,0.15,\ldots,0.40\}.$
The witness-validity tolerance is fixed at
$\varepsilon_{\mathrm{val}}={4}/{255}.$
Fresh base inputs are not reused within a
verification session.

For each session, $\mathcal{V}$ draws $N$ fresh
challenges according to this distribution, with base
inputs sampled without replacement within the
session:
\begin{equation}
C_i\sim\mathcal{D}_C,
\qquad
i=1,\ldots,N,
\end{equation}
where
\begin{equation}
C_i
=
(\bvec{x}_i,\psi_i)
\end{equation}
and
\begin{equation}
\psi_i
=
\left(
c'_i,
p^{\mathrm{target}}_{c'_i,i},
\varepsilon_{\mathrm{val}}
\right).
\end{equation}
$\bvec{x}_i, c'_i,$ and $p^\mathrm{target}_{c'_i,i}$
are not disclosed before the corresponding challenge
is issued.

The $\operatorname{ResGen}$ evaluated in
the main experiments is instantiated by the
Iterative Fine-grained Multi-class Probability
Control (I-FMPC) procedure, detailed in
Section~\ref{sec:ifmpc}.
Given $C_i$, $\mathcal{P}$ applies
$\operatorname{ResGen}=\operatorname{I\mathchar`-FMPC}$ to its local
model $M$:
\begin{equation}
\bvec{x}^{\mathrm{W}}_i
=
\operatorname{ResGen}(M,C_i)
=
\operatorname{I\mathchar`-FMPC}(M,C_i).
\label{eq:cnn_witness_generation}
\end{equation}
$\operatorname{ResGen}$ is used for both
$\Pi_{1}^{\mathcal{O}}$ and
$\Pi_{0}^{\mathcal{O}}$ in the reported main
experiments.

% The auxiliary class $c$ used internally by I-FMPC is
% not included in $\psi$.
% It is selected locally by $\mathcal{P}$ after the
% challenge is disclosed, as described in
% Section~\ref{sec:ifmpc}.

After receiving $\bvec{x}^{\mathrm{W}}_i$ from
$\mathcal{P}$, $\mathcal{V}$ first checks
\begin{equation}
\left\|\bvec{x}^{\mathrm{W}}_i-\bvec{x}_i\right\|_{\infty}
\leq
\varepsilon_{\mathrm{val}}.
\label{eq:witness_validity}
\end{equation}
Thus,
\begin{equation}
\operatorname{Valid}_{\mathrm{wit}}(C_i,\bvec{x}^{\mathrm{W}}_i)
=
\mathbb{I}
\left[
\left\|\bvec{x}^{\mathrm{W}}_i-\bvec{x}_i\right\|_{\infty}
\leq
\varepsilon_{\mathrm{val}}
\right].
\label{eq:cnn_witness_validity_function}
\end{equation}
The valid-response index set and count are
\begin{equation}
\begin{gathered}
\mathcal{I}_{\mathrm{val}}
=
\left\{
i\in\{1,\ldots,N\}
\,\middle|\,
\operatorname{Valid}_{\mathrm{wit}}(C_i,\bvec{x}^{\mathrm{W}}_i)=1
\right\},
\\
N_{\mathrm{val}}
=
\left|\mathcal{I}_{\mathrm{val}}\right|.
\end{gathered}
\label{eq:valid_response_set}
\end{equation}
If
$N_{\mathrm{val}}<N_{\min}$,
$\mathcal{V}$ rejects the session without computing
an aggregate score.
Otherwise, for each valid witness,
$\mathcal{V}$ obtains
\begin{equation}
\bvec{p}_i^{\mathrm{MLaaS}}
=
F^{\mathrm{MLaaS}}(\bvec{x}^{\mathrm{W}}_i)
\end{equation}
and defines the output probability of class $c'_i$ as
\begin{equation}
p^{\mathrm{MLaaS}}_{c'_i,i}
=
F^{\mathrm{MLaaS}}(\bvec{x}^{\mathrm{W}}_i)[c'_i].
\end{equation}
The per-challenge score is
\begin{equation}
E_i
=
D_i^{\mathrm{prob}}
=
\frac{
\left|
p^{\mathrm{target}}_{c'_i,i}
-
p^{\mathrm{MLaaS}}_{c'_i,i}
\right|
}{
p^{\mathrm{target}}_{c'_i,i}
}.
\label{eq:cnn_probability_discrepancy}
\end{equation}
For
$N_{\mathrm{val}}\geq N_{\min}$,
the aggregate score is
\begin{equation}
\Gamma_N
=
\overline{D}^{\mathrm{prob}}
=
\frac{1}{N_{\mathrm{val}}}
\sum_{i\in\mathcal{I}_{\mathrm{val}}}
D_i^{\mathrm{prob}}.
\label{eq:cnn_aggregate_score}
\end{equation}
The final decision rule is
\begin{equation}
\mathsf{Accept}
=
\begin{cases}
0,
&
N_{\mathrm{val}}<N_{\min},
\\
\mathbb{I}\!\left[
\overline{D}^{\mathrm{prob}}
\leq
\tau_{\mathcal{O}}
\right],
&
N_{\mathrm{val}}\geq N_{\min}.
\end{cases}
\label{eq:cnn_acceptance_rule}
\end{equation}
Invalid witnesses are excluded from the aggregate,
whereas $N_{\min}$ prevents acceptance based only on
a favorable subset of challenges.
$\tau_{\mathcal{O}}$ is calibrated
from reference configurations drawn from the
instantiated operating populations independently of
the configuration under evaluation.

Table~\ref{tab:framework_instantiation} summarizes the CNN instantiation of TP-CRIV.
\begin{table}[t]
\centering
\caption{Instantiation of the proposed TP-CRIV.}
\label{tab:framework_instantiation}
\begin{tabular}{c|l}
\hline
Component & Instantiation \\
\hline
$\mathcal{O}$
&
Population-scoped exact-instance distinction within
$\Pi^{\mathcal{O}}$
\\
$\operatorname{ResGen}$
&
Response-generation algorithm I-FMPC
\\
$\Pi_{1}^{\mathcal{O}}$
&
Same-model directed configurations in
$\mathcal{M}_{\mathrm{eval}}$
\\
$\Pi_{0}^{\mathcal{O}}$
&
Cross-model directed configurations in
$\mathcal{M}_{\mathrm{eval}}$
\\
$M$
&
Prover's CNN image classifier
\\
$M^{\mathrm{cloud}}$
&
Deployed MLaaS CNN image classifier
\\
$F^{\mathrm{MLaaS}}$
&
$F^{\mathrm{MLaaS}}(\bvec{x})
=
M^{\mathrm{cloud}}(\bvec{x})$
\\
$S$
&
Local input--probability geometry
\\
$N_{\min}$
&
$N$ in the reported CNN evaluation
\\
$\mathcal{D}_C$
&
ImageNet-based fresh probability-control challenges
\\
$\psi$
&
$(c',p^{\mathrm{target}}_{c'},
\varepsilon_{\mathrm{val}})$
\\
$W$
&
$\bvec{x}^{\mathrm{W}}$
\\
$\operatorname{Valid}_{\mathrm{wit}}$
&
$\mathbb{I}[\|W-\bvec{x}\|_{\infty}
\leq\varepsilon_{\mathrm{val}}]$
\\
$\operatorname{Eval}$
&
$D^{\mathrm{prob}}$
\\
$\operatorname{Aggr}$
&
Arithmetic mean over valid per-challenge scores
\\
$\Gamma_N$
&
$\overline{D}^{\mathrm{prob}}$
\\
$\mathsf{Accept}$
&
\makecell[l]{Reject if
$N_{\mathrm{val}}<N_{\min}$;\\
otherwise accept iff
$\overline{D}^{\mathrm{prob}}
\leq\tau_{\mathcal{O}}$}
\\
\hline
\end{tabular}
\end{table}

\subsection{Verification Principle}
\label{sec:cnn_verification_principle}
For the verification objective defined above, the selected model-dependent property should provide challenge-relevant information that induces distinguishable challenge-solving behavior between $\Pi_1^{\mathcal{O}}$ and $\Pi_0^{\mathcal{O}}$. In particular, under the instantiated challenge distribution and response-generation algorithm, configurations in $\Pi_1^{\mathcal{O}}$ should satisfy requirements determined by $M^{\mathrm{cloud}}$ more consistently than configurations in $\Pi_0^{\mathcal{O}}$.

For this purpose, we select $S$ as the local input--probability geometry, namely the relationship between bounded input modifications and the resulting class-probability changes. Behavior near decision boundaries has been used as model-sensitive information in prior fingerprinting studies \cite{IPGuard,AFA2020,LukasZK21,DeepFoolFP}, which motivates its use here as a property from which discriminative verification requirements can be constructed.

Based on this property, $\mathcal{V}$ issues fresh probability-control challenges. For a challenge
$C=(\bvec{x},c',p^{\mathrm{target}}_{c'},\varepsilon_{\mathrm{val}})$,
$\mathcal{P}$ must generate a witness satisfying
\begin{equation}
\left\|\bvec{x}^{\mathrm{W}}-\bvec{x}\right\|_{\infty}
\leq\varepsilon_{\mathrm{val}}
\end{equation}
and
\begin{equation}
F^{\mathrm{MLaaS}}(\bvec{x}^{\mathrm{W}})[c']
\approx p^{\mathrm{target}}_{c'}.
\label{eq:principle_probability_requirement}
\end{equation}
The corresponding solution region depends on $S_{M^{\mathrm{cloud}}}$, whereas $\mathcal{P}$ searches for a solution using $S_M$ through $\operatorname{ResGen}$. Varying $\bvec{x}$, $c'$, and $p^{\mathrm{target}}_{c'}$ therefore probes different local regions and probability levels.

For $(M,M^{\mathrm{cloud}})\in\Pi_1^{\mathcal{O}}$, the same model determines both the search and evaluation geometry. In contrast, for $(M,M^{\mathrm{cloud}})\in\Pi_0^{\mathcal{O}}$, a witness satisfying the target probability on $M$ need not lie in the corresponding solution region of $M^{\mathrm{cloud}}$. The procedure therefore induces the observable score distributions $P_{1,\operatorname{ResGen}}^{\mathcal{O}}$ and $P_{0,\operatorname{ResGen}}^{\mathcal{O}}$ under the instantiated $\mathcal{D}_C$ and $\operatorname{ResGen}$. The experiments below evaluate whether these distributions are sufficiently separated for threshold calibration and finite-challenge decisions.

\subsection{Witness Generation Algorithm: I-FMPC}
\label{sec:ifmpc}

For the $\operatorname{ResGen}$ characterized in the main experiments,
we use I-FMPC.
In the CNN-based instantiation of TP-CRIV, the
selected property $S$ is the local
input--probability geometry.
I-FMPC uses white-box information from $M$ to search
for a solution to each fresh probability-control
challenge.
Gradient information has been widely utilized in
white-box adversarial-example generation
\cite{FGSM,I-FGSM,madry2018towards}
as an efficient mechanism for identifying local
output-change directions.
I-FMPC uses this information to navigate the local
probability geometry of $M$ and locate a witness that
reaches the probability level designated by
$\mathcal{V}$ within the permitted neighborhood.

I-FMPC is designed for fine-grained probability
control rather than misclassification-oriented
optimization.
It adjusts each target-class probability toward a
specific value through adaptive balancing of the
target-class gradients and the gradient of an
auxiliary non-target class.
Let
\begin{equation}
\mathcal{T}'
=
\{c'_j\}_{j=1}^{m}
\subset
\mathcal{Y}
\end{equation}
denote the target-class set.
The auxiliary class $c$ is then selected as
\begin{equation}
c
=
\arg\max_{k\in\mathcal{Y}\setminus\mathcal{T}'}
M(\bvec{x})[k].
\label{eq:ifmpc_auxiliary_class}
\end{equation}
Thus, $c$ is the highest-probability class of $M$
among the classes outside $\mathcal{T}'$.
The controlled class set is then defined as
\begin{equation}
\mathcal{T}
=
\{c\}
\cup
\mathcal{T}'.
\end{equation}
For each $c'_j\in\mathcal{T}'$, $\mathcal{V}$
specifies a target probability
$p^{\mathrm{target}}_{c'_j}$.
We denote the set of target probabilities by
\begin{equation}
\bvec{p}^{\mathrm{target}}_{\mathcal{T}'}
=
\left\{
p^{\mathrm{target}}_{c'_j}
\right\}_{j=1}^{m}.
\end{equation}
The probabilities of the classes in $\mathcal{T}'$ are controlled toward these specified values. When these probabilities exceed their prescribed ranges, the contribution of the auxiliary class $c$ is increased to counterbalance further increases and help maintain a relatively high probability for $c$.

Given an input image $\bvec{x}$, I-FMPC generates a probability-controlled input $\bvec{x}^{\mathrm{W}} = \mathrm{I\mbox{-}FMPC}(M,C)$
with the primary objective
\begin{equation}
M(\bvec{x}^{\mathrm{W}})[c'_j] \approx p^{\mathrm{target}}_{c'_j},
\qquad j=1,\ldots,m. \label{eq:ifmpc_probability_goal}
\end{equation}
In addition, the generation algorithm encourages the probability of the auxiliary class $c$ to remain relatively high during probability control.
It should be noted that the auxiliary class $c$ is
not included in the verification requirement
specified by $\mathcal{V}$.
Instead, $c$ is selected and used solely by
$\mathcal{P}$ as an internal aid during witness
generation.

Let $p_k(\bvec{x})=M(\bvec{x})[k]$ denote the
softmax probability assigned to class $k$.
For each $k\in\mathcal{T}$, we define
\begin{equation}
L(\bvec{x},k)
=
-\log p_k(\bvec{x}).
\end{equation}
I-FMPC uses the weighted class-wise loss
\begin{equation}
L_{\mathrm{FMPC}}
\left(
\bvec{x}^{\mathrm{W}}_t
\right)
=
\sum_{k\in\mathcal{T}}
\beta^k
L
\left(
\bvec{x}^{\mathrm{W}}_t,k
\right),
\end{equation}
where $\beta^k$ controls the contribution of class $k$
to the combined input gradient.
The target-class terms increase the probabilities of
$c'_j\in\mathcal{T}'$, while the term for $c$
counterbalances their excessive increases.
The iterative update rule of I-FMPC is given by
\begin{equation}
\begin{gathered}
\bvec{x}^{\mathrm{W}}_0
=
\bvec{x},
\\[3pt]
\bvec{x}^{\mathrm{W}}_{t+1}
=
\operatorname{Clip}_{\bvec{x},\varepsilon_{\mathrm{gen}}}
\left\{
\bvec{x}^{\mathrm{W}}_t
-
\alpha^{\mathrm{com}}
\operatorname{sign}
\left(
\nabla_{\bvec{x}}
L_{\mathrm{FMPC}}
\left(
\bvec{x}^{\mathrm{W}}_t
\right)
\right)
\right\}
\\[3pt]
=
\operatorname{Clip}_{\bvec{x},\varepsilon_{\mathrm{gen}}}
\left\{
\bvec{x}^{\mathrm{W}}_t
-
\alpha^{\mathrm{com}}
\operatorname{sign}
\left(
\sum_{k\in\mathcal{T}}
\beta^k
\nabla_{\bvec{x}}
L
\left(
\bvec{x}^{\mathrm{W}}_t,
k
\right)
\right)
\right\}.
\end{gathered}
\label{eq:multi-class}
\end{equation}
Here, $\alpha^{\mathrm{com}}$ denotes the common update
step size, and $\beta^i$ controls the contribution of class
$i$ to the combined gradient.
It should be noted that the class-wise loss
$L(\bvec{x},i)$ does not directly contain the target
probability
$p^{\mathrm{target}}_{c'_j}$.
Instead, I-FMPC realizes fine-grained probability matching
through the adaptive control mechanism summarized in
Algorithm~\ref{alg:ifmpc}.
In the algorithm, $\operatorname{FMPC}(\cdot)$ denotes one
input-update step defined in Eq.~(\ref{eq:multi-class}).
For each target class $c'_j$, the corresponding weight
$\beta^{c'_j}$ is increased when its recent average
probability remains below the prescribed target range.
Conversely, when at least one target-class probability
exceeds its prescribed range while all target-class
probabilities remain above their lower bounds, the weight
$\beta^c$ of the auxiliary non-target class is increased
to counterbalance the excessive increase in the
target-class probabilities and help maintain a relatively
high probability for $c$.
Once all target-class probabilities enter the neighborhood
of their specified values, the common step size
$\alpha^{\mathrm{com}}$ is reduced for finer convergence.
Thus, I-FMPC achieves probability matching by adaptively
balancing the class-wise cross-entropy gradients and refining
the update step near the target region, rather than by directly
minimizing a probability-distance loss.
\begin{algorithm}[t]
\caption{I-FMPC: Iterative Fine-grained Multi-class Probability Control}
\label{alg:ifmpc}
\begin{algorithmic}[1]
\Require
Input image $\bvec{x}$,
model $M$ with output-label space $\mathcal{Y}$,
target class set
$\mathcal{T}'=\{c'_j\}_{j=1}^{m}$,
target probabilities
$\bvec{p}^{\mathrm{target}}_{\mathcal{T}'}
=
\{p^{\mathrm{target}}_{c'_j}\}_{j=1}^{m}$,
shared step size $\alpha^{\mathrm{com}}$,
generation perturbation budget
$\varepsilon_{\mathrm{gen}}$,
averaging interval $l$,
tolerance $T^{\mathrm{diff}}$,
maximum iterations $t^{\max}$,
minimum step threshold $\alpha_{\mathrm{th}}$,
step decay factor $\gamma\in(0,1)$
\Ensure
Probability-controlled input
$\bvec{x}^{\mathrm{W}}$
\State
$\bvec{x}^{\mathrm{W}}_{0}=\bvec{x}$
\State
$c
\gets
\arg\max_{k\in\mathcal{Y}\setminus\mathcal{T}'}
M(\bvec{x})[k]$
\State
$\mathcal{T}\gets\{c\}\cup\mathcal{T}'$
\State
$\bvec{x}^{\mathrm{W}}
=
\bvec{x}^{\mathrm{W}}_{0}$
\State
$\beta^{i}
\gets
1,
\quad
\forall i\in\mathcal{T}$
\For{$t=1$ to $t^{\max}$}

    \State
    $\bvec{x}^{\mathrm{W}}_{t}
    =
    \operatorname{FMPC}
    \left(
    \bvec{x}^{\mathrm{W}}_{t-1},
    \bvec{x},
    M,
    \alpha^{\mathrm{com}},
    \{\beta^{i}\}_{i\in\mathcal{T}},
    \varepsilon_{\mathrm{gen}}
    \right)$

    \State
    $\bvec{x}^{\mathrm{W}}
    \gets
    \bvec{x}^{\mathrm{W}}_{t}$

    \State
    $\bvec{p}_{t}
    =
    M(\bvec{x}^{\mathrm{W}}_{t})$

    \If{$t \bmod l = 0$}

        \State
        $\bvec{p}^{\mathrm{mean}}
        =
        \dfrac{1}{l}
        \sum_{k=t-l+1}^{t}
        \bvec{p}_{k}$

        \For{each $j=1$ to $m$}
            \If{$
            p^{\mathrm{mean}}_{c'_j}
            <
            (1-T^{\mathrm{diff}})
            p^{\mathrm{target}}_{c'_j}
            $}
                \State
                $\beta^{c'_j}
                \gets
                \beta^{c'_j}+1$
            \EndIf
        \EndFor

        \If{
        $\left(
        \exists j\in\{1,\dots,m\}:
        p^{\mathrm{mean}}_{c'_j}
        >
        (1+T^{\mathrm{diff}})
        p^{\mathrm{target}}_{c'_j}
        \right)$
        \textbf{and}
        $\left(
        \forall j\in\{1,\dots,m\}:
        p^{\mathrm{mean}}_{c'_j}
        >
        (1-T^{\mathrm{diff}})
        p^{\mathrm{target}}_{c'_j}
        \right)$
        }
            \State
            $\beta^{c}
            \gets
            \beta^{c}+1$
        \EndIf

        \If{
        $
        \displaystyle
        \max_{j\in\{1,\dots,m\}}
        \left|
        \frac{
        p^{\mathrm{mean}}_{c'_j}
        -
        p^{\mathrm{target}}_{c'_j}
        }{
        p^{\mathrm{target}}_{c'_j}
        }
        \right|
        \le
        T^{\mathrm{diff}}
        $
        }
            \State
            $\alpha^{\mathrm{com}}
            \gets
            \gamma
            \alpha^{\mathrm{com}}$
        \EndIf

        \If{$\alpha^{\mathrm{com}}<\alpha_{\mathrm{th}}$}
            \State
            \textbf{break}
        \EndIf

    \EndIf

\EndFor

\State
\Return
$\bvec{x}^{\mathrm{W}}$
\end{algorithmic}
\end{algorithm}

\begin{table}[t]
\centering
\caption{
I-FMPC settings used for the representative
probability-control trajectories in
Fig.~\ref{fig:I-FMPC}.
}
\label{tab:ifmgsm_params}
\begin{tabular}{c|cccc}
\hline
 & $T^{\mathrm{diff}}$ & $\varepsilon_{\mathrm{gen}}$ & $\alpha^{\mathrm{com}}$ & $\bvec{p}^{\mathrm{target}}_{\bvec{c'}}$ \\
\hline
$m=1$ & $1\!\times\!10^{-3}$ & $4/255$ & $1\!\times\!10^{-3}$ & $\{0.30\}$ \\
$m=2$ & $1\!\times\!10^{-3}$ & $12/255$ & $5\!\times\!10^{-3}$ & $\{0.30,\, 0.15\}$ \\
$m=3$ & $3\!\times\!10^{-3}$ & $12/255$ & $5\!\times\!10^{-3}$ & $\{0.30,\, 0.15,\, 0.10\}$ \\
\hline
\end{tabular}
\end{table}
\begin{table}[t]
\centering
\caption{Common hyperparameter settings for I-FMPC.}
\label{tab:ifmpc_common_params}
\begin{tabular}{c|c}
\hline
Hyperparameter & Setting \\
\hline
Maximum iterations $t^{\max}$ & $10{,}000$ \\
Averaging interval $l$ & $5$ \\
Minimum step threshold $\alpha_{\mathrm{th}}$ & $1\times10^{-10}$ \\
Step decay factor $\gamma$ & $0.5$ \\
\hline
\end{tabular}
\end{table}
\begin{figure}[t]
\centering
\begin{minipage}[b]{0.7\columnwidth}
    \includegraphics[width=\textwidth]{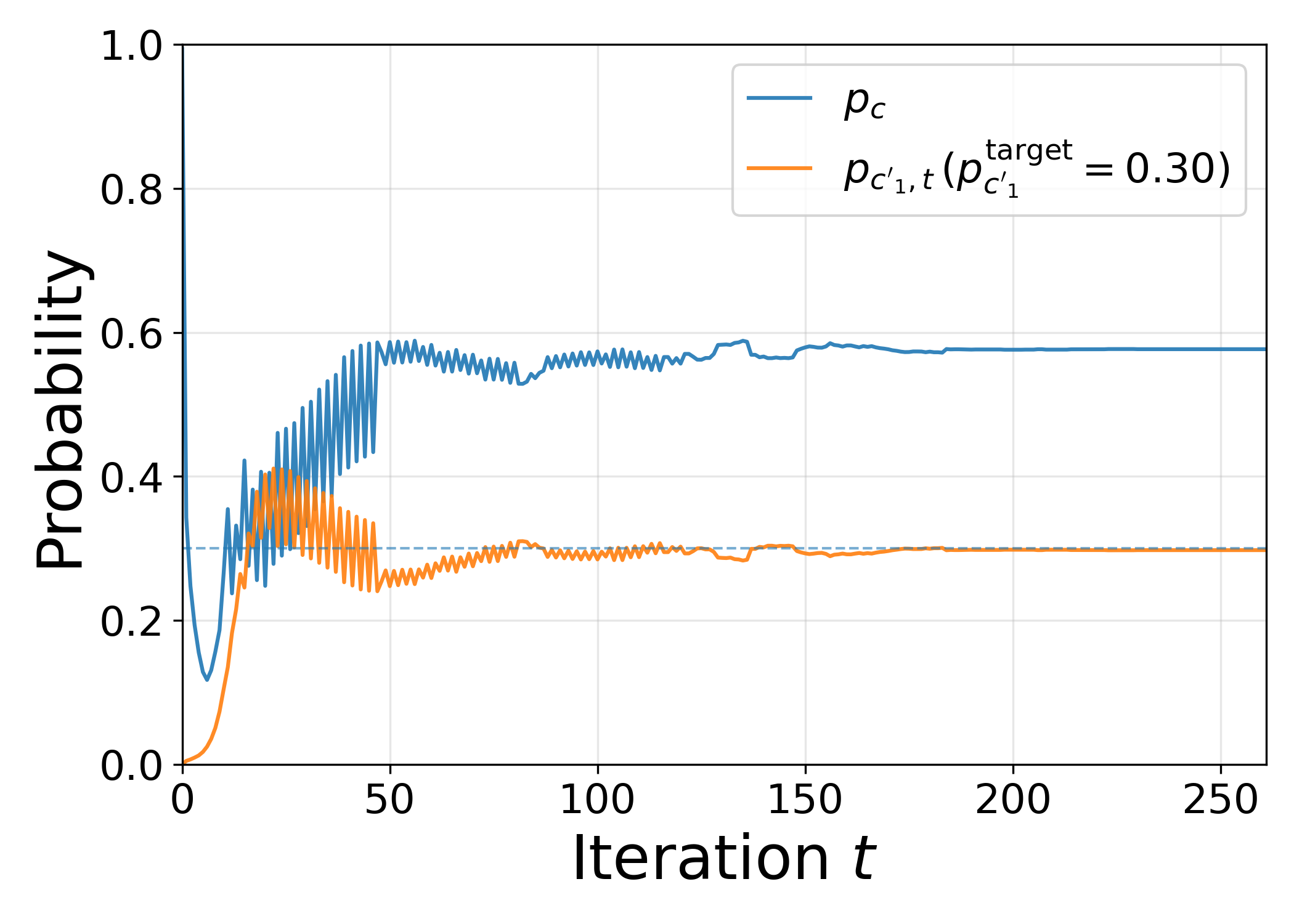}
    \smallskip
    \centering\footnotesize\text{(a) $m=1$}
\end{minipage}
\begin{minipage}[b]{0.7\columnwidth}
    \includegraphics[width=\textwidth]{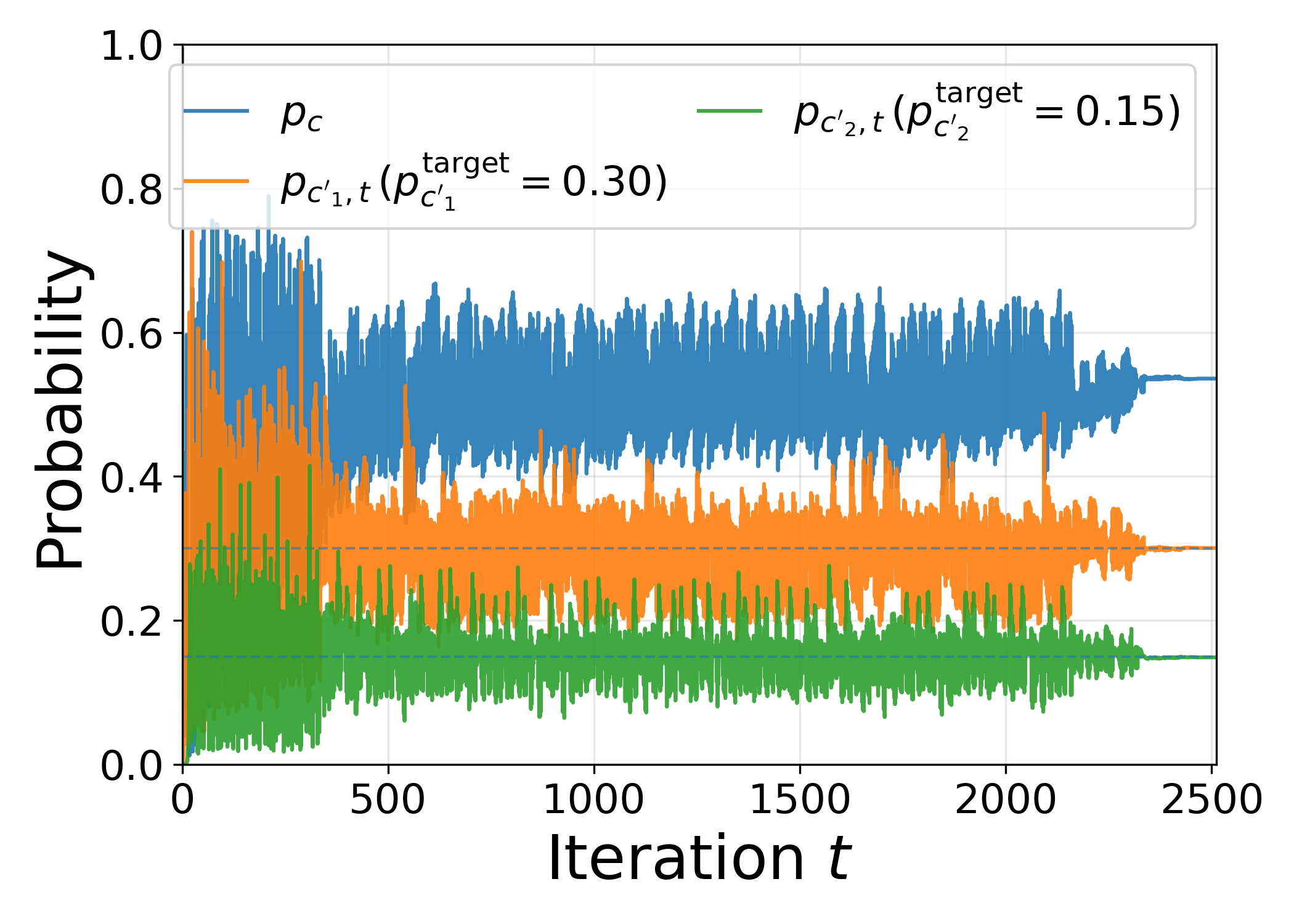}
    \smallskip
    \centering\footnotesize\text{(b) $m=2$}
\end{minipage}
\begin{minipage}[b]{0.7\columnwidth}
    \includegraphics[width=\textwidth]{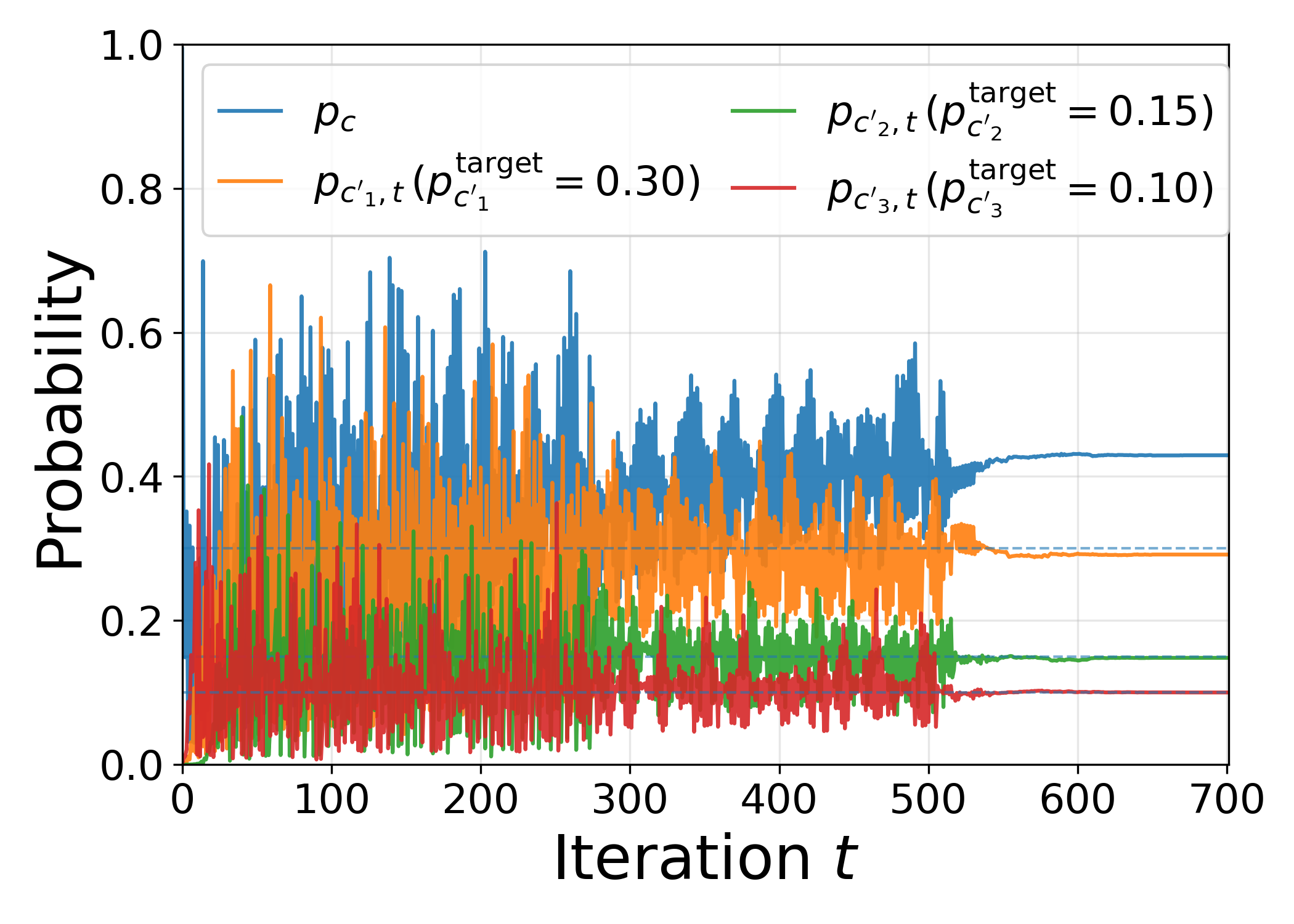}
    \smallskip
    \centering\footnotesize\text{(c) $m=3$}
\end{minipage}
\caption{
Trajectory of output probabilities
$p_{k,t}$
for the controlled classes
$k\in\mathcal{T}$
over iterations $t$, where
$\bvec{x}^{\mathrm{W}}_t$
denotes the image generated by I-FMPC at iteration
$t$ and
$p_{k,t}
=
M(\bvec{x}^{\mathrm{W}}_t)[k]$.
The probabilities of the target classes in $\mathcal{T}'$ converge toward their designated values, while the auxiliary class $c$ selected by $\mathcal{P}$ maintains a relatively high probability throughout the illustrated trajectories.
}
\label{fig:I-FMPC}
\end{figure}

In the experiments presented in this paper, we
employ the dual-class configuration
($m=1$).
For each target class $c'$ specified by
$\mathcal{V}$, $\mathcal{P}$ selects
\begin{equation}
c
=
\arg\max_{k\in\mathcal{Y}\setminus\{c'\}}
M(\bvec{x})[k]
\end{equation}
as the auxiliary competing class.
I-FMPC then controls the probability of $c'$ toward
$p^{\mathrm{target}}_{c'}$ while using $c$ to
counterbalance excessive changes in the target-class
probability.
% The generated probability-controlled input serves
% as the witness returned by $\mathcal{P}$ in response
% to a verifier-generated challenge.
% Consequently, I-FMPC functions as the practical
% realization of the witness-generation algorithm
% $\operatorname{ResGen}$ in the proposed framework.
To illustrate the behavior of I-FMPC,
Fig.~\ref{fig:I-FMPC} presents representative
probability trajectories for
$m=1,2,$ and $3$.
The corresponding configuration-dependent
hyperparameter settings are summarized in
Table~\ref{tab:ifmgsm_params}, while the common
hyperparameter settings used throughout the
evaluations are listed in
Table~\ref{tab:ifmpc_common_params}.
As shown in Fig.~\ref{fig:I-FMPC}, the target-class
probabilities approach their specified values, while
the auxiliary class provides a counterbalancing contribution during
probability control.

By utilizing gradient information derived from $M$,
I-FMPC enables $\mathcal{P}$ to search for a witness
satisfying the fresh requirement $\psi$ specified
by $\mathcal{V}$.
The algorithm therefore realizes $\operatorname{ResGen}$ by using
$S_M$ to construct a candidate solution.
Whether this candidate also satisfies the requirement
on $M^{\mathrm{cloud}}$ is then evaluated through
$F^{\mathrm{MLaaS}}$.
Thus, I-FMPC operationalizes the central TP-CRIV
principle: information about the selected property is
not compared directly, but is used to solve a
challenge whose valid solution is determined by that
property.

\subsection{Feasibility Assessment}
\label{sec:requirement_assessment}

This subsection empirically characterizes the
challenge-induced score behavior of the instantiated
operating populations
$\Pi_{1}^{\mathcal{O}}$ and
$\Pi_{0}^{\mathcal{O}}$
under $\mathcal{D}_C$ and
$\operatorname{ResGen}$ specified
above.
Specifically, we examine whether I-FMPC can solve the
property-demanding challenges for configurations in
$\Pi_{1}^{\mathcal{O}}$, whether the resulting
scores are separated from those obtained for
configurations in $\Pi_{0}^{\mathcal{O}}$, and
whether a threshold estimated from calibration
configurations transfers to held-out configurations
within the same operating world.

\subsubsection{Experimental Setting}

The evaluation includes all 10 configurations in
$\Pi_{1}^{\mathcal{O}}$ and all 90 configurations
in $\Pi_{0}^{\mathcal{O}}$ constructed from
$\mathcal{M}_{\mathrm{eval}}$.
For both $\Pi_{1}^{\mathcal{O}}$ and
$\Pi_{0}^{\mathcal{O}}$,
$\operatorname{ResGen}=\operatorname{I\mathchar`-FMPC}$ is fixed.
Importantly, the non-matching configurations use the
same fine-grained probability-control algorithm and
optimize the same target probability specified by
$\mathcal{V}$ on their respective models.

$\mathcal{P}$ uses the generation budget
\begin{equation}
\varepsilon_{\mathrm{gen}}
=
\varepsilon_{\mathrm{val}}
=
\frac{4}{255}.
\end{equation}
Witnesses are generated using the $m=1$
configuration of I-FMPC.
The configuration-dependent parameters are reported
in Table~\ref{tab:ifmgsm_params}, while the shared
stopping and step-control parameters are listed in
Table~\ref{tab:ifmpc_common_params}.
The target probability specified by $\mathcal{V}$ is sampled
per challenge as described in the verification
procedure above.

All witnesses generated in the model-pair and
threshold-calibration experiments satisfied
\begin{equation}
\left\|
\bvec{x}^{\mathrm{W}}-\bvec{x}
\right\|_{\infty}
\leq
\varepsilon_{\mathrm{gen}}
\leq
\varepsilon_{\mathrm{val}},
\end{equation}
and I-FMPC reached its stopping condition before the
maximum iteration limit.
Therefore, all returned witnesses were valid and
included in the subsequent score evaluation.
Consequently,
$N_{\mathrm{val}}=N$
for every verification trial reported below.
For this CNN instantiation, we set
$N_{\min}=N,$
so a session cannot be accepted by returning valid
witnesses only for a favorable subset of the issued
challenges.
To further characterize the generated witnesses, we
measured their perceptual similarity to the
corresponding base images.
Across the ten evaluated models, the mean SSIM \cite{SSIM}
ranged from $0.9890$ to $0.9933$, the mean LPIPS \cite{Zhang2018LPIPS}
from $0.0024$ to $0.0056$, and the mean RMSE from
$0.0049$ to $0.0060$.
These results indicate that the probability-control
requirements are satisfied using only small and
perceptually subtle modifications to the base
images.

\subsubsection{Observable Separation Induced by Property-Demanding Challenges}

To characterize the score behavior induced by
$\Pi_{1}^{\mathcal{O}}$ and
$\Pi_{0}^{\mathcal{O}}$, we generate
100 fresh challenges sampled according to
$\mathcal{D}_C$ for each directed configuration in
$\Pi_{1}^{\mathcal{O}}
\cup
\Pi_{0}^{\mathcal{O}}.$
For every challenge,
$\operatorname{ResGen}=\operatorname{I\mathchar`-FMPC}$ is applied to
$M$.
Thus, matching and non-matching configurations use
the same probability-control objective and algorithm;
only the relationship between the local and deployed
models changes.

\begin{figure}[t]
    \centering
    \includegraphics[width=0.9\linewidth]
    {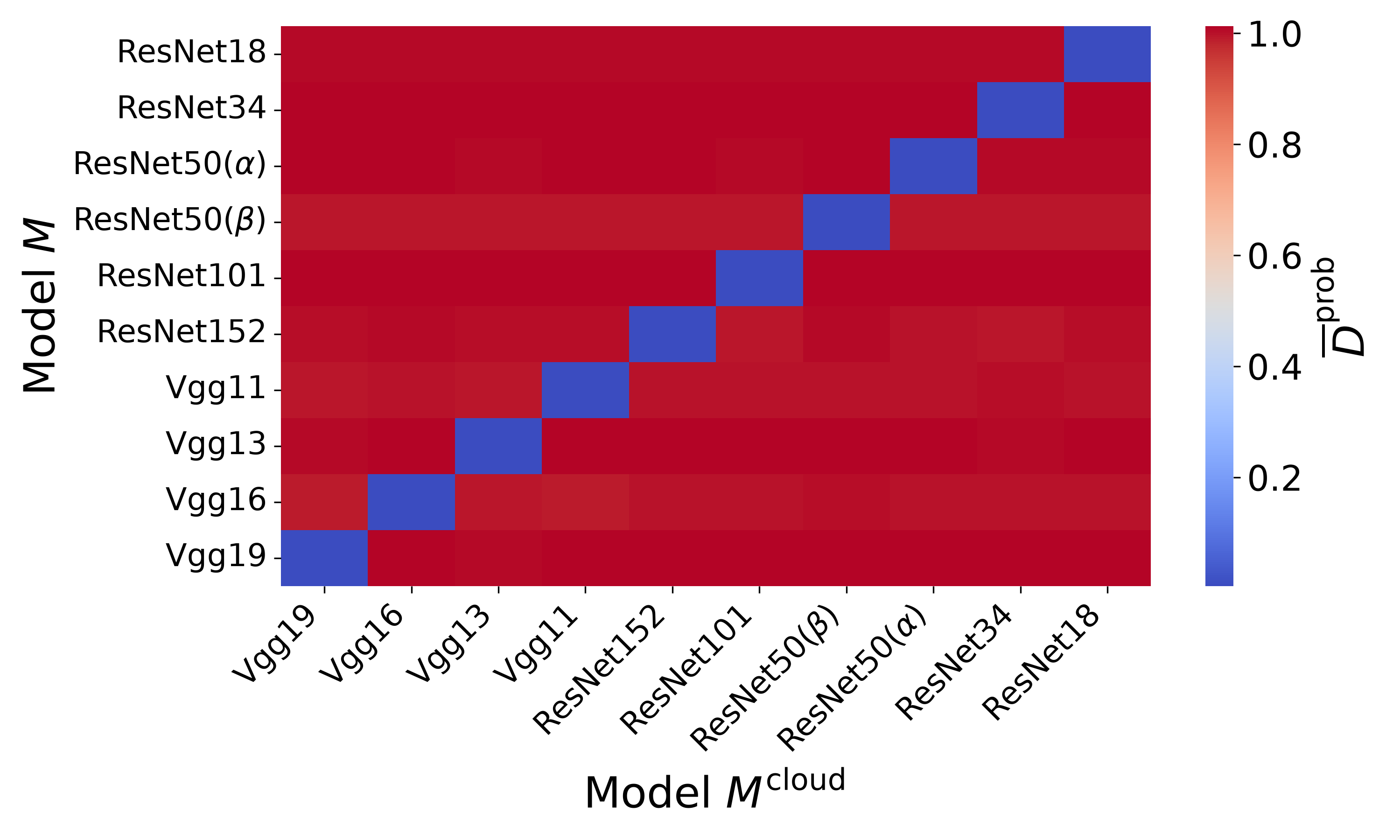}
\caption{
Heatmap of
$\overline{D}^{\mathrm{prob}}$
for the evaluated directed model configurations.
The diagonal entries correspond to same-model
configurations, whereas the off-diagonal entries
correspond to cross-model configurations.
}
    \label{ex:Dprob_clean}
\end{figure}

For each directed model configuration, the
per-challenge discrepancies are aggregated as
\begin{equation}
\Gamma_{100}
=
\overline{D}^{\mathrm{prob}}
=
\frac{1}{100}
\sum_{i=1}^{100}
D_i^{\mathrm{prob}}.
\end{equation}
Fig.~\ref{ex:Dprob_clean} shows the resulting
$\overline{D}^{\mathrm{prob}}$ for all evaluated
configurations.

For configurations in
$\Pi_{1}^{\mathcal{O}}$, represented by the
diagonal entries, $\overline{D}^{\mathrm{prob}}$
remains close to zero.
Because $M=M^{\mathrm{cloud}}$, I-FMPC uses the same
local probability information that determines the
challenge solution on $M^{\mathrm{cloud}}$.
These results therefore demonstrate that the
selected challenge family is consistently solvable
for $\Pi_{1}^{\mathcal{O}}$ under the specified
$\operatorname{ResGen}$.

By contrast, all configurations in
$\Pi_{0}^{\mathcal{O}}$, represented by the
off-diagonal entries, exhibit substantially larger
values of $\overline{D}^{\mathrm{prob}}$.
For these configurations, I-FMPC locates a solution
with respect to the local input--probability geometry of
$M$, whereas the requirement is evaluated on
$M^{\mathrm{cloud}}$.
The resulting witness therefore generally fails to
reach the probability designated by $\mathcal{V}$ on
$M^{\mathrm{cloud}}$.

Thus, under the instantiated $\mathcal{D}_C$ and
$\operatorname{ResGen}$, the evaluated
property-demanding challenge family
induces observable separation between the score
behavior of
$\Pi_{1}^{\mathcal{O}}$ and
$\Pi_{0}^{\mathcal{O}}$.
The heatmap does not represent a direct distance
between complete model decision landscapes.
It represents finite-challenge solving error produced
by the declared challenge-response procedure.

The 100-challenge experiment characterizes
separation within $\Pi^{\mathcal{O}}$ only.
The next experiment examines whether this separation
supports binary decisions with fewer fresh
challenges.

\subsubsection{Finite-Challenge Verification and Threshold Calibration}

The preceding experiment shows observable separation
between the score behavior induced by
$\Pi_{1}^{\mathcal{O}}$ and
$\Pi_{0}^{\mathcal{O}}$ under the instantiated $\mathcal{D}_C$ and $\operatorname{ResGen}$.
We next examine whether this separation can be
converted into finite-challenge binary decisions
using an operational threshold calibrated
independently of the configuration under evaluation.

To emulate threshold determination by $\mathcal{V}$,
the ten models in
$\mathcal{M}_{\mathrm{eval}}$
are divided into calibration and held-out model sets.
For each calibration setting, three models are
selected as calibration models held by $\mathcal{V}$, while
the remaining seven models are reserved for held-out
evaluation.
All
$\binom{10}{3}=120$
possible selections of the three calibration models
are evaluated.
For each selection, the same-model directed
configurations induced by the three calibration
models form the calibration subset of
$\Pi_{1}^{\mathcal{O}}$, whereas the cross-model
directed configurations among those models form the
calibration subset of
$\Pi_{0}^{\mathcal{O}}$.
The corresponding configurations induced by the
seven remaining models are used for held-out
evaluation.
Thus, calibration and held-out evaluation use
disjoint model subsets while preserving the same population-scoped identity
objective, $\mathcal{D}_C$, $\operatorname{ResGen}$, and decision rule.

The 100 available challenge samples are randomly
divided into two disjoint sets of 50 calibration
challenges and 50 test challenges.
The calibration challenges are used only for
threshold determination, whereas the test challenges
are used only for held-out finite-challenge
evaluation.
For threshold calibration, we fix the calibration
aggregation size to
$N_{\mathrm{cal}}=10.$
For each calibration-model configuration,
non-overlapping groups of
$N_{\mathrm{cal}}$
calibration challenges are formed, and the
corresponding aggregate score is
\begin{equation}
\Gamma_{N_{\mathrm{cal}}}
=
\frac{1}{N_{\mathrm{cal}}}
\sum_{i=1}^{N_{\mathrm{cal}}}
D_i^{\mathrm{prob}}.
\label{eq:cnn_calibration_aggregate}
\end{equation}
The matching and non-matching calibration
configurations therefore induce empirical
distributions of
$\Gamma_{N_{\mathrm{cal}}}$
using only models held by $\mathcal{V}$ and calibration
challenges.

$\tau_{\mathcal{O}}$
is selected as
\begin{equation}
\tau_{\mathcal{O}}
=
\frac{
\max
\Gamma_{N_{\mathrm{cal}}}^{\mathrm{same}}
+
\min
\Gamma_{N_{\mathrm{cal}}}^{\mathrm{cross}}
}{2},
\label{eq:cnn_threshold_calibration}
\end{equation}
where the maximum same-model score and minimum
cross-model score are computed using only the
calibration configurations and calibration
challenges.
If the calibration distributions are separated such that
$\max \Gamma_{N_{\mathrm{cal}}}^{\mathrm{same}}
<
\min \Gamma_{N_{\mathrm{cal}}}^{\mathrm{cross}}$,
$\tau_{\mathcal{O}}$ is selected as
the midpoint between these two values.
Otherwise,
$\tau_{\mathcal{O}}$
is selected as the operating point that minimizes
the absolute difference between the
false-acceptance and false-rejection rates on the
calibration data.

Importantly, in this experiment the resulting
$\tau_{\mathcal{O}}$
is determined once for each calibration-model
selection and is intentionally kept fixed as the
number of verification challenges changes.

For held-out evaluation, we consider
$N\in\{1,3,5,10\}.$
For each evaluated value of $N$, non-overlapping
groups of $N$ test challenges are used to construct
one verification trial.
With 50 held-out test challenges, this yields
50, 16, 10, and 5 complete groups per directed model
configuration for $N=1,3,5,$ and $10$,
respectively.
For $N=3$, the two remaining challenges are not used
in a complete group.
$\mathcal{V}$ computes
\begin{equation}
\Gamma_N
=
\frac{1}{N}
\sum_{i=1}^{N}
D_i^{\mathrm{prob}}.
\label{eq:cnn_finite_challenge_aggregate}
\end{equation}
Because
$N_{\mathrm{val}}=N_{\min}=N$
for every trial in this evaluation, the general
acceptance rule in
Eq.~\eqref{eq:cnn_acceptance_rule}
reduces to
\begin{equation}
\mathsf{Accept}=1
\quad\Longleftrightarrow\quad
\Gamma_N
\leq
\tau_{\mathcal{O}}.
\label{eq:cnn_finite_challenge_acceptance}
\end{equation}
Accordingly, the same
$\tau_{\mathcal{O}}$
is applied to
$\Gamma_1$,
$\Gamma_3$,
$\Gamma_5$, and
$\Gamma_{10}$
within each calibration-model selection.

For statistical reporting, one verification trial is
defined as one value of $\Gamma_N$ obtained from one
directed model configuration and one non-overlapping
group of $N$ held-out test challenges under one fixed
calibration-model selection.
Acceptance of a configuration in
$\Pi_{0}^{\mathcal{O}}$
is counted as a false acceptance, whereas rejection
of a configuration in
$\Pi_{1}^{\mathcal{O}}$
is counted as a false rejection.
Trials constructed from the same directed model
configuration are not regarded as independent
model-level samples.
For each of the 120 calibration-model selections,
FAR, FRR, and AUC are computed separately using the
corresponding seven held-out models and test
challenges.

Across the 120 calibration-model selections, the
fixed threshold calibrated using
$N_{\mathrm{cal}}=10$
has a mean of
$0.4868$,
a standard deviation of
$0.0114$,
and a range from
$0.4653$ to $0.5035$.
This variation reflects the choice of the three
calibration models held by $\mathcal{V}$.

Table~\ref{tab:threshold_calibration} reports the
held-out finite-challenge results obtained using
these fixed thresholds.
FAR, FRR, and AUC are computed separately for each
of the 120 calibration-model selections.
Because the 120 calibration-model selections overlap, we report the distribution of FAR and FRR across selections.
\begin{table}[t]
\centering
\caption{
Held-out finite-challenge verification using a fixed
operational threshold calibrated with
$N_{\mathrm{cal}}=10$.
For each of the 120 calibration-model selections,
the resulting
$\tau_{\mathcal{O}}$
is applied unchanged to all evaluated values of
$N$.
The reported FAR and FRR are the median and range
over the 120 selections.
}
\label{tab:threshold_calibration}
\begin{tabular}{c c c}
\hline
$N$ &
\begin{tabular}{c}
Error rate across selections \\
median [min, max]
\end{tabular}
&
Min. AUC
\\
\hline

1 &
\begin{tabular}{c}
FAR: 0 [0, 0]\% \\
FRR: 0 [0, 0]\%
\end{tabular}
&
1.000
\\
\hline

3 &
\begin{tabular}{c}
FAR: 0 [0, 0]\% \\
FRR: 0 [0, 0]\%
\end{tabular}
&
1.000
\\
\hline

5 &
\begin{tabular}{c}
FAR: 0 [0, 0]\% \\
FRR: 0 [0, 0]\%
\end{tabular}
&
1.000
\\
\hline

10 &
\begin{tabular}{c}
FAR: 0 [0, 0]\% \\
FRR: 0 [0, 0]\%
\end{tabular}
&
1.000
\\
\hline
\end{tabular}
\end{table}
No false acceptance or false rejection was observed
for any of the 120 calibration-model selections or
for any evaluated value of
$N\in\{1,3,5,10\}$.
Accordingly, the median, minimum, and maximum of the
split-specific FAR and FRR are all zero.
The minimum split-specific AUC is also 1.000 for
every evaluated value of $N$.
Thus, even for $N=1$, the fixed threshold calibrated
from
$\Gamma_{10}$
on the calibration models held by $\mathcal{V}$ separated
all evaluated matching and non-matching held-out
trials.

In the present experiment, the observable separation
is sufficiently large that perfect held-out
classification is retained for all evaluated values
of $N$.
However the 120 calibration-model selections should not be interpreted as 120 independent
experimental replications.
Because the selections overlap, each same-model
directed configuration appears in the held-out set
for
$\binom{9}{3}=84$
selections, whereas each cross-model directed
configuration appears in the held-out set for
$\binom{8}{3}=56$
selections.
The same challenge instances are also reused across
different calibration-model selections.
The split-specific results therefore constitute an
exhaustive sensitivity analysis with respect to the
choice of calibration models rather than independent
estimates from randomly repeated experiments.

At the distinct-configuration level, the held-out
evaluations collectively cover all 10 elements of
$\Pi_{1}^{\mathcal{O}}$ and all 90 elements of
$\Pi_{0}^{\mathcal{O}}$.
None of the 90 non-matching configurations was
accepted under any calibration-model selection in
which that configuration was held out.
Similarly, none of the 10 matching configurations
was rejected.
Thus, the observed zero-error result is shared by
all distinct configurations in the instantiated
operating populations and is not produced solely by
pooling repeated challenge-level observations.

These results indicate that, within the instantiated
operating world, the observable separation
characterized in Fig.~\ref{ex:Dprob_clean} can be
converted into finite-challenge decisions using a
fixed operational threshold estimated from separate
models held by $\mathcal{V}$.
Accordingly, when $\mathcal{V}$ observes an accepted
verification result for a suspicious configuration,
the demonstrated challenge-solving performance is
interpreted as being more consistent with
$\Pi_{1}^{\mathcal{O}}$ than with
$\Pi_{0}^{\mathcal{O}}$ within the calibrated scope.
Together with freshness, network isolation, and the
model-based response-generation assumption, this
provides an empirical basis for inferring that
$\mathcal{P}$ locally possesses a model satisfying
the declared identity relative to
$M^{\mathrm{cloud}}$.

The absence of false rejection provides empirical
evidence of finite-challenge completeness for
$\Pi_{1}^{\mathcal{O}}$ under the evaluated
response-generation setting, whereas the absence of
false acceptance provides empirical evidence of
soundness-related behavior for
$\Pi_{0}^{\mathcal{O}}$ under the same condition.
The reported empirical FAR and FRR summarize the
observed finite-challenge error behavior of the
evaluated configurations under the instantiated
$\mathcal{D}_C$, $\operatorname{ResGen}$, and deployment pipeline.
They should not be interpreted as population-level
guarantees for model configurations or
response-generation algorithms outside this
experimental scope.
The model configurations also share constituent
models and should not be regarded as independent
draws from a broader population of AI models.
Accordingly, the results do not provide a universal
threshold guarantee, a broader population-level
confidence interval, or proof that every
non-matching model must fail the selected challenge
family.
\subsubsection{Assessment of the Protocol Requirements}

The evaluated CNN instantiation satisfies the
abstract requirements within the declared
experimental scope as follows.

\paragraph{Completeness}
For $\Pi_{1}^{\mathcal{O}}$,
$M=M^{\mathrm{cloud}}$ and
$\operatorname{ResGen}=\operatorname{I\mathchar`-FMPC}$.
Table~\ref{tab:threshold_calibration} reports no
held-out rejection for any evaluated $N$.
These observations provide empirical evidence of
finite-challenge completeness for
$\Pi_{1}^{\mathcal{O}}$.

\paragraph{Soundness}
For $\Pi_{0}^{\mathcal{O}}$, the same $\operatorname{ResGen}$ searches
using $S_M$ while the requirement is evaluated using
$M^{\mathrm{cloud}}$.
Table~\ref{tab:threshold_calibration} reports no
false acceptance for any evaluated $N$.
This soundness-related evidence applies only to the
90 declared non-matching configurations and the fixed
$\operatorname{ResGen}$ used in the experiment.

\paragraph{Property-Demanding Challenge Design}
Each $C_i$ varies $\bvec{x}_i$, $c'_i$, and
$p^{\mathrm{target}}_{c'_i,i}$, and each valid
$W_i$ must remain within
$\varepsilon_{\mathrm{val}}$ of $\bvec{x}_i$.
Fig.~\ref{ex:Dprob_clean} shows that, for configurations in $\Pi_1^{\mathcal{O}}$, the resulting witnesses consistently satisfy the specified fine-grained probability requirements on $M^{\mathrm{cloud}}$, whereas for configurations in $\Pi_0^{\mathcal{O}}$, the witnesses generally fail to transfer the same requirements to $M^{\mathrm{cloud}}$.

\paragraph{Witness Validity}
All reported witnesses satisfy
$\varepsilon_{\mathrm{gen}}\leq
\varepsilon_{\mathrm{val}}$, yielding
$N_{\mathrm{val}}=N$ in every reported trial.
With $N_{\min}=N$, acceptance therefore cannot be
obtained by returning valid witnesses only for a
favorable subset of challenges.

\paragraph{Freshness and Network Isolation}
Base inputs are not reused within a session and the
remaining challenge components are freshly sampled
from $\mathcal{D}_C$.
The operational interpretation additionally assumes
the network-isolation condition of
Section~\ref{sec:framework}.
Together with the calibrated matching decision,
these conditions satisfy the freshness and
network-isolation requirement within the evaluated
model-based scope.

\section{Concluding Remarks}
\label{sec:conclusion}
This paper proposed TP-CRIV, a challenge-response
framework for third-party identity verification of
AI models deployed through remote services.
TP-CRIV considers a setting in which the verifier
$\mathcal{V}$ has neither white-box nor API access
to the claimant's model $M$, interacts with the
suspicious deployed model $M^{\mathrm{cloud}}$ only
through its ordinary black-box service interface,
and requires no protocol-specific cooperation from
the service provider.
Under these conditions, TP-CRIV enables
$\mathcal{V}$ to obtain empirical evidence as to
whether the prover $\mathcal{P}$ locally possesses a
model satisfying the identity specified by the
intended verification objective $\mathcal{O}$.

For each $\mathcal{O}$, matching and non-matching
operating situations,
$\Pi_1^{\mathcal{O}}$ and $\Pi_0^{\mathcal{O}}$,
are specified a priori.
A model-dependent property $S$ is then used to
construct fresh property-demanding challenges, and
the resulting challenge-solving performance is
interpreted using calibration on independent
reference configurations.
Within the model-based response-generation scope,
freshness and network isolation exclude online
external assistance after challenge disclosure.
Matching-consistent performance therefore provides
an empirical basis for inferring that
the prover $\mathcal{P}$ possesses a model satisfying the
identity declared by $\mathcal{O}$.
This inference is statistical and does not establish
training provenance or legal ownership.

We instantiated TP-CRIV using ten
ImageNet-pretrained CNN models with
$\operatorname{ResGen}=\mathrm{I\mathchar`-FMPC}$.
Within the evaluated operating population
$\Pi^{\mathcal{O}}$, same-model configurations were
assigned to $\Pi_1^{\mathcal{O}}$, whereas
cross-model configurations were assigned to
$\Pi_0^{\mathcal{O}}$.
The two situations exhibited clear score separation,
and independently calibrated thresholds produced no
observed false acceptance or false rejection for
$N\in\{1,3,5,10\}$ across all 120 calibration-model
selections.
These results demonstrate feasibility within the
explicitly evaluated operating scope rather than
universal soundness over arbitrary models or
response-generation mechanisms.

Several directions remain for extending the
framework and its operational guarantees.

\begin{itemize}

\item
\textbf{Broader operating populations:}
Future implementations should extend both
$\Pi_1^{\mathcal{O}}$ and $\Pi_0^{\mathcal{O}}$
according to the intended verification objective
$\mathcal{O}$.
These populations may include independently trained,
fine-tuned, pruned, quantized, extracted, distilled,
and otherwise transformed models, with each
configuration assigned to the matching or
non-matching situation according to the identity
semantics of $\mathcal{O}$.
For example, a transformed model may belong to
$\Pi_0^{\mathcal{O}}$ under an exact-instance
objective but to $\Pi_1^{\mathcal{O}}$ under a
lineage-oriented objective.
Broader populations would enable more reliable
characterization of FAR, FRR, threshold transfer,
and soundness within the declared operating scope.

\item
\textbf{Broader non-matching prover strategies:}
Future work should consider a broader range of
response-generation strategies $\operatorname{ResGen}$ available to
provers whose model configurations belong to
$\Pi_0^{\mathcal{O}}$.
Even for the same non-matching configuration,
stronger optimization, ensemble-based strategies,
or other model-based procedures may change
challenge-solving performance and false-acceptance
behavior.
A further extension is to consider non-model
response mechanisms, such as learned
challenge-to-witness generators, which lie outside
the present model-based response-generation and
soundness scope.

\item
\textbf{Challenge design and deployment conditions:}
Future work should investigate challenge-selection
strategies, the relationship between challenge
diversity and the number of challenges $N$,
practical enforcement or auditing of network
isolation, and robustness to service-side
preprocessing, model modification, and stochastic
inference.

\item
\textbf{Broader AI-model instantiations:}
Future work should instantiate and evaluate TP-CRIV
for a wider range of AI model types, including
generative, language, multimodal, and other AI
systems.
For each model type, suitable model-dependent
properties $S_M$, property-demanding challenge
families, and response-generation algorithms $\operatorname{ResGen}$
should be identified and evaluated according to the
intended verification objective $\mathcal{O}$ and
operating scope.

\end{itemize}

\bibliographystyle{ieeetr}
\bibliography{refs}

\end{document}